\documentclass[12pt,letterpaper]{article}
\usepackage[a4paper, total={7in, 10in}]{geometry}

\usepackage{graphicx}
\usepackage{helvet}
\usepackage{authblk}
\usepackage[hidelinks]{hyperref}
\usepackage{amsmath} 
\usepackage{amssymb} 
\usepackage{setspace}
\usepackage{comment}
\usepackage{cleveref}
\usepackage{xcolor,colortbl}
\usepackage{tikz}
\usepackage{float}
\usepackage{booktabs}
\usepackage{multirow}
\usepackage{lscape}
\usepackage[separate-uncertainty = true,multi-part-units=single]{siunitx}
\DeclareSIUnit{\pixel}{px}
\DeclareSIUnit{\fov}{fov}
\DeclareSIUnit{\fps}{fps}
\DeclareSIUnit{\rpm}{rpm}
\DeclareSIUnit{\bar}{bar}

\usepackage{orcidlink} 
\usepackage[super,comma,sort&compress]  
   {natbib}

\usepackage{tikz}
\definecolor{cultivationchamber}{RGB}{128,128,224}

\usepackage{xcolor}                                                       
\newcommand{\bluerect}{\fcolorbox{blue}{white}{\rule{0pt}{0.35em}\rule{0.35em}{0pt}}}                               
\newcommand{\tealrect}{\fcolorbox{teal}{white}{\rule{0pt}{0.35em}\rule{0.35em}{0pt}}}   

\usepackage[acronym]{glossaries}
\makeglossaries
\newacronym{cad}{CAD}{Computer Aided Design}
\newacronym{pdms}{PDMS}{Polydimethylsiloxane}
\newacronym{tsca}{TSCA}{Total Single-Cell Area}
\newacronym{dart}{DART}{Design-Aware and Real-Time capable}
\newacronym{sak}{SAK}{Swiss Army Knife}
\newacronym{sam}{SAM}{Segment Anything Model}
\newacronym{yolo}{YOLO}{You Only Look Once}
\newacronym{fps}{FPS}{Frames per Second}
\newacronym{cif}{CIF}{Caltech Intermediate Format}
\newacronym{fov}{FoV}{Field of View}
\newacronym{cglut}{\textit{C.~glutamicum}}{\textit{Corynebacterium glutamicum}}
\newacronym[longplural={Regions of Interest}]{roi}{RoI}{Region of Interest}
\newacronym[longplural={Identifiers}]{id}{ID}{Identifier}
\newacronym{mlci}{MLCI}{Microfluidic Live-Cell Imaging}

\newcounter{sisection}

\crefname{sisection}{}{}
\Crefname{sisection}{}{}
\newif\ifsifirst
\sifirsttrue

\makeatletter
\renewcommand{\maketitle}{\bgroup\setlength{\parindent}{0pt}
\begin{flushleft}
  \textbf{\@title}
  
  \@author
\end{flushleft}\egroup}
\makeatother

\title{DART: A design-aware microfluidic chip paradigm for real-time live-cell image analysis}
\date{}

\author[1,\orcidlink{0000-0002-2087-9847}]{Johannes Seiffarth} 
\author[1,2,\orcidlink{0000-0001-8260-1245}]{Matthias Pesch} 
\author[3,\orcidlink{0000-0002-3959-733X}]{Lukas Scholtes}
\author[1,\orcidlink{0000-0001-6215-1857}]{Dietrich Kohlheyer} 
\author[3,\orcidlink{0000-0002-8555-6416}]{Hanno Scharr}
\author[1,\orcidlink{0000-0002-5407-2275}*]{Katharina Nöh} 

\affil[1]{Institute for Bio- and Geosciences, IBG-1: Biotechnology, Forschungszentrum Jülich, Jülich, Germany}
\affil[2]{Computational Systems Biotechnology (AVT.CSB), RWTH Aachen University, Aachen, Germany}
\affil[3]{Institute for Advanced Simulation, IAS-8: Data Analytics and Machine Learning, Forschungszentrum Jülich, Jülich, Germany}
\affil[*]{Correspondence: k.noeh@fz-juelich.de}

\begin{document}

\maketitle

\section*{SUMMARY}

High-throughput microfluidic live-cell imaging generates rich single-cell data. Yet semi-automated procedures for locating regions of interest (RoIs), each containing one cell population, and removing surrounding microfluidic structures from recorded images, scale with the number of RoIs. This prevents real-time image analysis and delays time-to-insight by hours to days. We introduce the Design-Aware and Real-Time capable (DART) paradigm for microfluidic cultivation chips, which aligns the CAD blueprint with the physical chip and thereby enables throughput-independent localization of all RoIs and fully automated image processing across diverse RoI geometries and chip layouts. DART establishes this alignment through embedded fiducial markers and deep-learning-based marker detection. We validate DART using the Swiss Army Knife chip, which combines eight structurally distinct RoI designs across \num{1164} RoI locations. DART localizes all RoIs in five minutes, removes microfluidic structures from raw microscopy images in \SI{40}{\milli\second}, and performs fully automated image analysis, including cell segmentation, in under \SI{1.1}{\second} per image. Together, these capabilities establish DART as an end-to-end hardware-software paradigm with real-time-capable analysis that paves the way toward closed-loop and outcome-driven smart microscopy.

\section*{KEYWORDS}

Microfluidic live-cell imaging, Design-aware image analysis, High-throughput cultivation, Smart microscopy, CAD blueprint-guided analysis, Fiducial marker registration, Real-time image processing, Single-cell segmentation, Swiss Army Knife chip design, Open-source platform

\section*{INTRODUCTION}

Microfluidic live-cell imaging has become a powerful platform for studying microbial cell behavior at the single-cell level. Picoliter-scale bioreactors enable long-term cultivation and observation of individual bacteria under precisely controlled conditions~\cite{wang_robust_2010,grunberger_disposable_2012,grunberger_microfluidic_2013}, while advanced microfluidic chip designs provide dynamic environmental control, including oxygen gradients~\cite{Kasahara2025}, and enable monitoring of gene expression dynamics~\cite{kaiser_monitoring_2018}. Modern microfluidic chips host hundreds to thousands of \glspl{roi}, each containing a single cell population, thereby enabling concurrent observation of many cell populations at single-cell resolution~\cite{grunberger_spatiotemporal_2015,helfrich_live_2015,sachs_image_2016}. Converting these high-throughput image data into quantitative single-cell measurements, however, requires more than cell segmentation alone: it also requires identifying the cultivation area and distinguishing it from surrounding microfluidic structures.

\begin{figure*}
    \centering
    \includegraphics[width=0.9\linewidth]{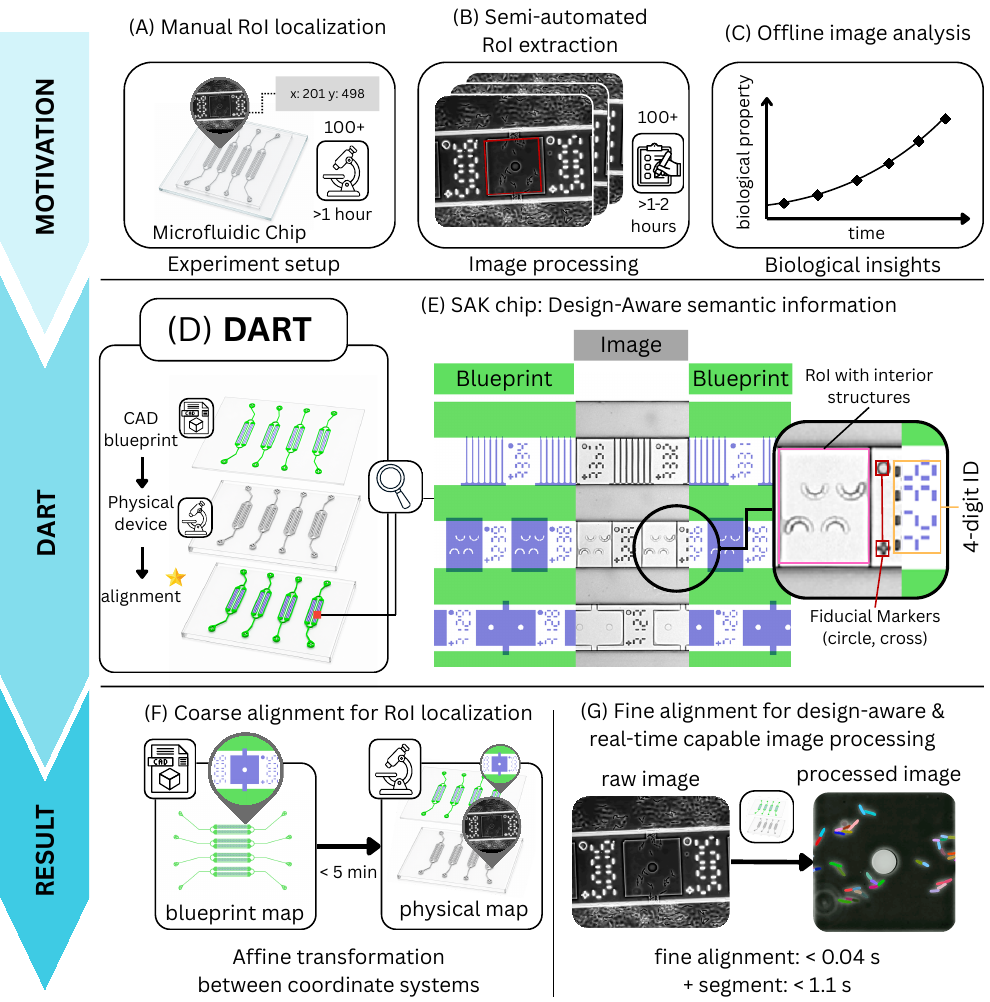}
    \caption{\textbf{Overview of the \gls{dart} microfluidic cultivation platform.} High-throughput live-cell imaging is limited by (A) time-consuming manual \gls{roi} localization, (B) semi-automated image processing, and (C) offline image analysis that delays quantitative readouts such as single-cell or population development. (D) The \gls{dart} paradigm addresses these limitations by aligning the \gls{cad} blueprint with the physical microfluidic chip, thereby linking device design to the real device. This establishes (E) structural awareness of each \gls{roi} and its surrounding microfluidic features. \gls{dart} performs two levels of alignment: (F) a coarse blueprint-to-device alignment for rapid, throughput-independent \gls{roi} localization in microscope stage coordinates, and (G) a fine blueprint-to-image alignment for fully automated, real-time-capable image processing, including masking of interior structures (gray) and cell segmentation (colored instances) with sub-micron precision. See also Supplemental Movie~S1.}
    \label{fig:overview}
\end{figure*}

In practice, however, most \gls{mlci} workflows remain dominated by manual steps (\Cref{fig:overview}A-C). Before each experiment, operators must locate every \gls{roi} on the microscope stage because chip mounting introduces an unknown offset between the design blueprint and the stage coordinate system. As a result, setup effort scales with the number of imaged \glspl{roi}, often consuming hours for high-throughput experiments~\cite{blobaum_protocol_2023} (\Cref{fig:overview}A). This throughput-dependent setup burden limits the practical scalability of modern high-density chip designs.

A second bottleneck arises during image analysis: Because microfluidic cultivations are performed in structurally diverse \glspl{roi}, recorded images typically contain not only the cultivation area in which living cells grow, but also surrounding microfluidic structures and, in some cases, cells outside the relevant cultivation region such as in media supply channels (\Cref{fig:overview}B). Robust quantitative analysis therefore requires separating cells inside the \gls{roi} from both microfluidic structures and irrelevant surrounding areas. This is particularly challenging in devices with interior features such as mother-machine traps~\cite{oconnor_delta_2022}, cell-trapping structures~\cite{prangemeier_yeast_2022}, or supporting pillars~\cite{zhou_computer_2023}. Although modern segmentation methods~\cite{cutler_omnipose_2022,stringer_cellpose_2021,pachitariu_cellpose-sam_2025,marks_cellsam_2025,archit_microsam_2024} accurately detect cells, they generally do not encode the microfluidic context of the image and therefore cannot determine whether segmented objects belong to the relevant cultivation region. Consequently, masking of microfluidic structures and exclusion of cells outside the \gls{roi} are often performed semi-automatically after the experiment~\cite{blobaum_protocol_2023,kasahara_unveiling_2025,witting_microfluidic_2025,dal_co_short-range_2020,grunberger_spatiotemporal_2015,helfrich_live_2015}. Like the manual \glspl{roi} setup, this preprocessing burden scales with throughput~\cite{thiermann_tools_2024,dal_co_short-range_2020} and typically delays quantitative insight by days after data acquisition (\Cref{fig:overview}C).

To address these challenges, several automated image-analysis approaches have been developed~\cite{oconnor_delta_2022,thiermann_tools_2024,kaiser_monitoring_2018}. Methods such as DeLTA have demonstrated the potential of real-time image analysis for closed-loop microscopy~\cite{lugagne_deep_2024}. However, existing approaches typically remain tied to specific microfluidic chamber geometries, such as mother-machine devices~\cite{thiermann_tools_2024}. Extending them to new \gls{roi} architectures usually requires new training data, manual annotation, and retraining of the underlying image-analysis models~\cite{oconnor_delta_2022}. Thus, despite major progress in real-time segmentation, the structural knowledge already available in the chip design is generally not used during the experiment itself hampering full analysis automation.

Here, we introduce the \gls{dart} paradigm for microfluidic cultivation chips, which links a chip's \gls{cad} blueprint -- containing the positions and structural features of all \glspl{roi} -- to its physical realization on the microscope stage~(\Cref{fig:overview}D-E). To establish this link, fiducial markers and \glspl{id} are embedded directly into the chip design next to each \gls{roi}. \gls{dart} detects these markers in microscopy images and uses them to align blueprint and device by an affine transformation~(\Cref{fig:overview}E). This design-aware linkage enables two capabilities: first, automatic localization of all \glspl{roi} through a fixed-effort coarse blueprint-to-device alignment procedure (\Cref{fig:overview}F); and second, fine blueprint-to-image alignment for fully automated masking of cultivation structures during image analysis (\Cref{fig:overview}G). Because the relevant structural information is taken directly from the blueprint rather than inferred separately for each geometry, \gls{dart} transfers across \gls{roi} positions and across the major microfluidic \gls{roi} design classes currently used in the field, including layouts with complex interior structures. A video of the \gls{dart} image-processing workflow across the eight \gls{sak} \glspl{roi} designs is provided in Supplemental Movie~S1.

To validate this concept, we developed the \gls{sak} chip as a high-throughput \gls{dart} implementation that combines eight established \gls{roi} architectures on a single device~\cite{grunberger_spatiotemporal_2015,merrin_frontiers_2019,long_microfluidic_2013,wang_robust_2010}. We further developed open-source software for fiducial detection, coarse alignment for \gls{roi} localization, and fine alignment for real-time-capable image processing. Using this combined hardware-software platform, we validate throughput-independent \gls{roi} localization, fully automated image processing, and end-to-end analysis in an exemplary \gls{cglut} cultivation experiment. Together, these results establish \gls{dart} as a design-aware framework for scalable and automated microfluidic live-cell imaging.

\section*{RESULTS}

\subsection*{The DART paradigm and Swiss Army Knife Chip}

\begin{figure}
    \centering
    \includegraphics[width=.7\linewidth]{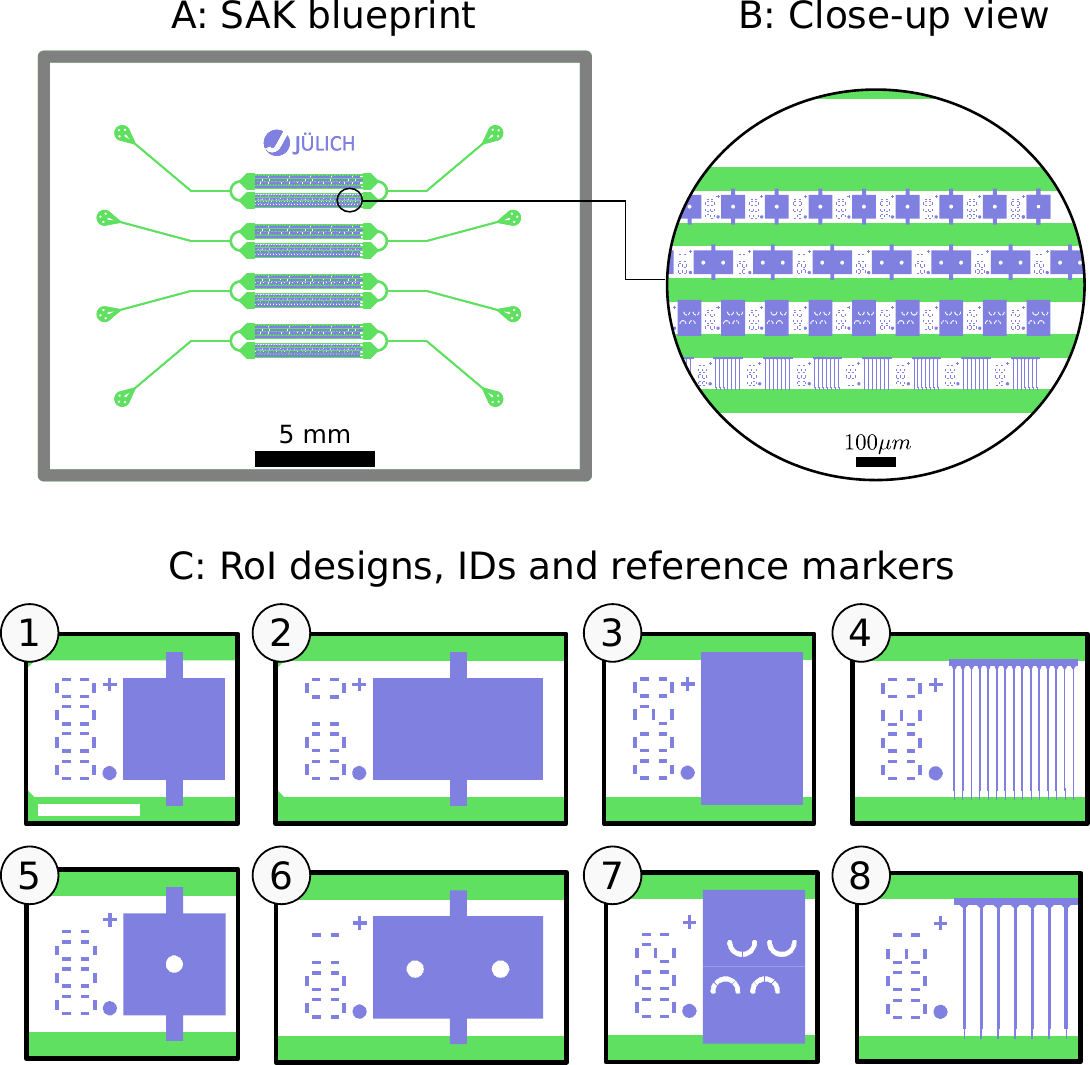}
    \caption[short]{\textbf{\gls{cad} layout of the \gls{sak} chip and its \gls{roi} architectures.} (A)~Overview of the full chip, including the inlets and outlets of the four independently supplied media channels. (B)~Close-up of the grid layout within a single channel. \glspl{roi} are arranged in rows and columns; the first two digits of the unique \gls{id} encode the row position and the last two digits encode the column position. Each \gls{roi} is further equipped with two fiducial markers (\textcolor{cultivationchamber}{\textbf{$+$}} and \tikz\draw[cultivationchamber,fill=cultivationchamber] (0,0) circle (.5ex);). (C)~Detailed views of the eight different \gls{roi} designs integrated in the \gls{sak}. The white scale bar in (C1) denotes \SI{60}{\micro\metre} and applies to C1-C8. Colors indicate fabrication heights of \SI{10}{\micro\metre} (green) and \SI{1}{\micro\metre} (purple).}
    \label{fig:sak_blueprint}
\end{figure}

To validate \gls{dart} across structurally diverse microfluidic cultivation geometries, we designed the \gls{sak} chip as a high-throughput platform that combines multiple established \gls{roi} architectures on a single device. Beyond serving as a validation platform for \gls{dart}, this design also enables side-by-side comparison of cultivation performance across diverse micro-environments within one experiment. The \gls{sak} follows common high-throughput microfluidic layouts~\cite{grunberger_disposable_2012,wang_robust_2010}: it comprises four separate medium channels (\Cref{fig:sak_blueprint}A, green), each containing a grid of \glspl{roi} arranged in rows and columns (\Cref{fig:sak_blueprint}A,B, purple).

In contrast to conventional high-throughput microfluidic chips that typically use a single \gls{roi} design, the \gls{sak} integrates eight distinct \gls{roi} architectures, arranged such that each row per medium channel contains one design class (\Cref{fig:sak_blueprint}A,C). These \glspl{roi} were selected to represent major microfluidic cultivation geometries currently used in the field~\cite{grunberger_spatiotemporal_2015,oconnor_delta_2022} and span a range of sizes (\Cref{SItab:roi_dimensions}). In addition, the \gls{sak} includes \gls{roi} designs with interior microfluidic features to assess \gls{dart}'s ability to remove structures in immediate proximity to living cells (e.g., \Cref{fig:sak_blueprint}C~5-7).

Following the \gls{dart} paradigm, each \gls{roi} on the \gls{sak} is equipped with two fiducial markers (a cross and a circle) and a unique \gls{id} that encodes its grid position (\Cref{fig:sak_blueprint}C and Methods). The first two digits denote the row position and the last two digits the column position. Together, these design elements establish an unambiguous link between each physical \gls{roi} and its corresponding location in the \gls{cad} blueprint. The human-readable \gls{id} also facilitates manual navigation at high magnification and small \glspl{fov} by allowing operators to directly reference their position in the blueprint.

In total, the \gls{sak} contains \num{1164} \glspl{roi} and was fabricated using established microfluidic chip manufacturing methods (Methods). This makes the \gls{sak} both a high-throughput cultivation platform and a rigorous test-bed for evaluating \gls{dart} across the major \gls{roi} architectures currently used for microbial \gls{mlci}.

\subsection*{Real-time fiducial marker detection}

Fiducial marker detection is the technical prerequisite for both coarse blueprint-to-device alignment and fine blueprint-to-image alignment in the \gls{dart} pipeline. It therefore must combine high positional accuracy with low inference latency. To identify a suitable detector, we benchmarked models from the \gls{yolo} object-detection family, which is widely used for real-time image analysis~\cite{shi_smartllsm_2024,waithe_object_2020,alhamadani_yolo_deepsort_2025}. We recorded and annotated a dataset of \num{319} microscopy images containing fiducial markers (\Cref{SIfig:train-batch}), split it into training, validation, and test sets, and trained multiple \gls{yolo} models using the Ultralytics framework~\cite{jocher_ultralytics_2023}. We benchmarked \gls{yolo} versions, model sizes, input resolutions, and detection tasks to select the detector used in the \gls{dart} pipeline, while applying microscopy-motivated data augmentation during training (see Methods). Full benchmark results are provided in Supplemental Text~\Cref{sec:SI_yolo_benchmark} and \Cref{SItab:yolo_results}, and a benchmark summary is shown in \Cref{SIfig:benchmark_marker_center_error_and_speed}.

The best-performing model (\gls{yolo}v26-s, \SI{1280}{\pixel}, detection task) achieved a \num{95}th-percentile marker-center detection error of \SI{0.18}{\micro\metre} (\SI{2.81}{\pixel}) with an inference time of \SI[separate-uncertainty=true]{16.4(1.0)}{\milli\second} per image on the benchmarking hardware (see Methods). This combination of sub-micron localization accuracy and millisecond-scale inference enables reliable fiducial marker detection for both coarse and fine alignment in the \gls{dart} pipeline and enables real-time image analysis.

\subsection*{Coarse alignment: Throughput-independent \gls{roi} localization}

\gls{dart} uses fiducial marker detection to compute a coarse alignment between the \gls{cad} blueprint and the physical microfluidic chip on the microscope stage. This alignment yields the stage coordinates of all \glspl{roi} from the blueprint and thereby replaces throughput-scaling manual \gls{roi} localization during experiment setup.

\begin{figure}
    \centering
    \includegraphics[width=.9\linewidth]{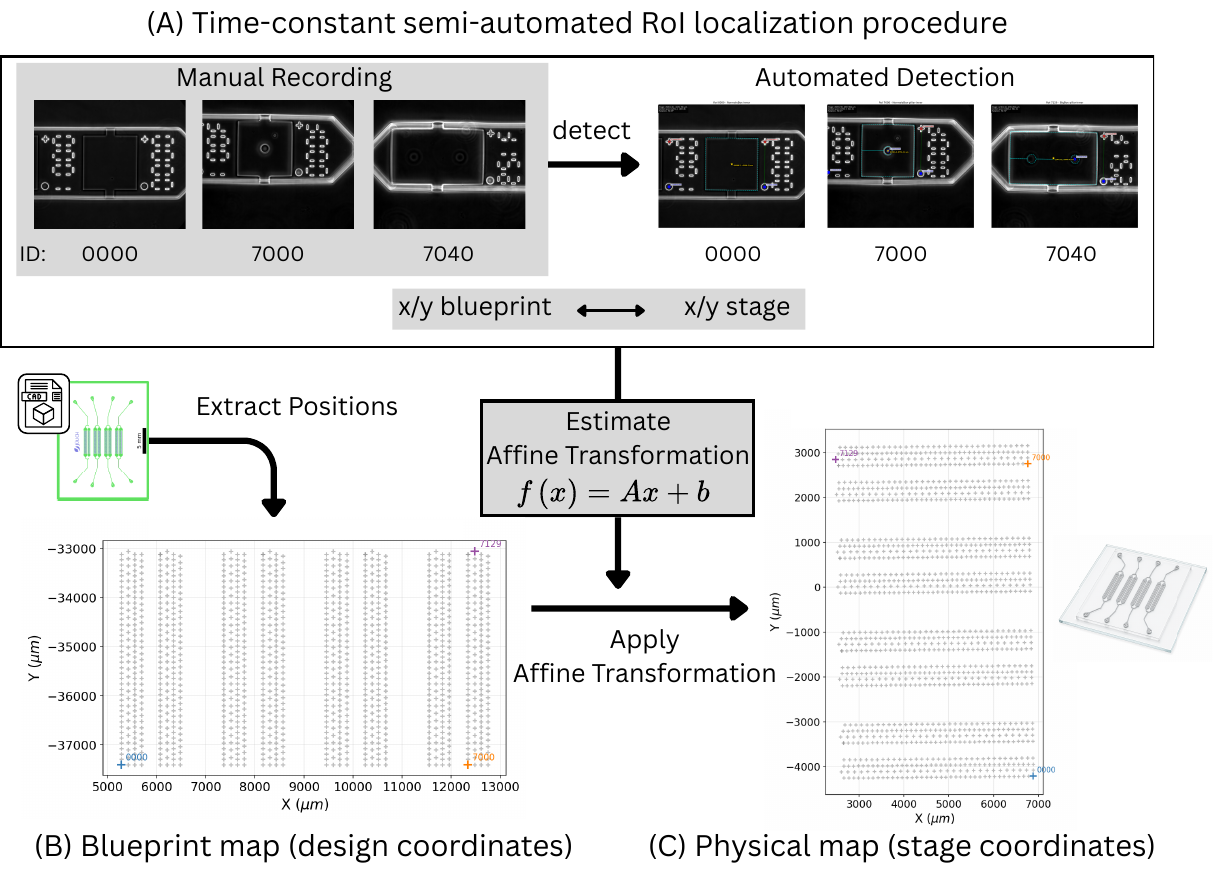}
    \caption{\textbf{Aligning \gls{cad} blueprint and microscope stage coordinate systems.} (A)~The operator visits three \glspl{roi} and records an image, the corresponding \gls{roi} ID, and the microscope stage position. \gls{dart} detects the fiducial markers, determines the \gls{roi} centers, and computes an affine transformation from the corresponding coordinate pairs in blueprint and stage coordinates. This transformation maps the \gls{cad} blueprint (B) onto the physical chip on the microscope stage (C) and localizes all \glspl{roi}. 
    Alignment accuracy in terms of localization errors and their spatial distribution across the microfluidic device are shown in \Cref{SIfig:benchmark_marker_center_error_and_speed}~A-B.}   
    \label{fig:dart_mapping_workflow}
\end{figure}

The coarse alignment workflow is outlined in \Cref{fig:dart_mapping_workflow}A and requires roughly five minutes of operator time. The operator manually visits three \glspl{roi} using the microscope stage positioning controls, records a phase-contrast image for each \gls{roi}, notes its \gls{id}, and stores the corresponding stage position. These three reference \glspl{roi} should be positioned at the edge or the corner of the chip to increase robustness against localization errors. 
\gls{dart} then detects the fiducial markers, determines the \gls{roi} center positions in stage coordinates, and computes an affine transformation from the corresponding \gls{roi} positions in blueprint and stage coordinates (Text~\Cref{sec:SI_map_creation}). This transformation maps all blueprint \gls{roi} coordinates into the microscope stage coordinate system (\Cref{fig:dart_mapping_workflow}B,C), making the full chip layout directly accessible for automated navigation. Because this mapping is derived from the blueprint, the localization procedure is independent of the total number of \glspl{roi} and readily transferable across different chip layouts.

We evaluated the localization accuracy of the coarse alignment on the \gls{sak}. After computing the affine transformation, we compared the predicted and experimentally observed center positions of individual \glspl{roi} in stage coordinates and quantified the resulting L2 localization error (see Methods). 
The spatial distribution of the localization errors across the full \gls{sak} chip and their summary distribution is shown in \Cref{SIfig:dmc_map_validation}~A-B.
The median L2 localization error was \SI{10.46}{\micro\metre}, with a \num{90}th percentile of \SI{18.39}{\micro\metre}. Relative to the microscope's \gls{fov} of \num{168.4} $\times$ \SI{142.1}{\micro\metre\squared}, these errors correspond to approximately \SI{6.4}{\percent} and \SI{17.6}{\percent} of the \gls{fov} width, respectively. 
These localization errors likely arise from a combination of microscope stage inaccuracy, affine transformation error, fabrication tolerances, and deformation of the soft \gls{pdms} chip material. Representative examples illustrate localization errors caused by structural deformation, fine feature geometry, and fabrication defects (\Cref{SIfig:dmc_map_validation}).

\subsection*{Fine alignment: Design-aware real-time image processing}

\begin{figure}
    \centering
    \includegraphics[width=.9\linewidth]{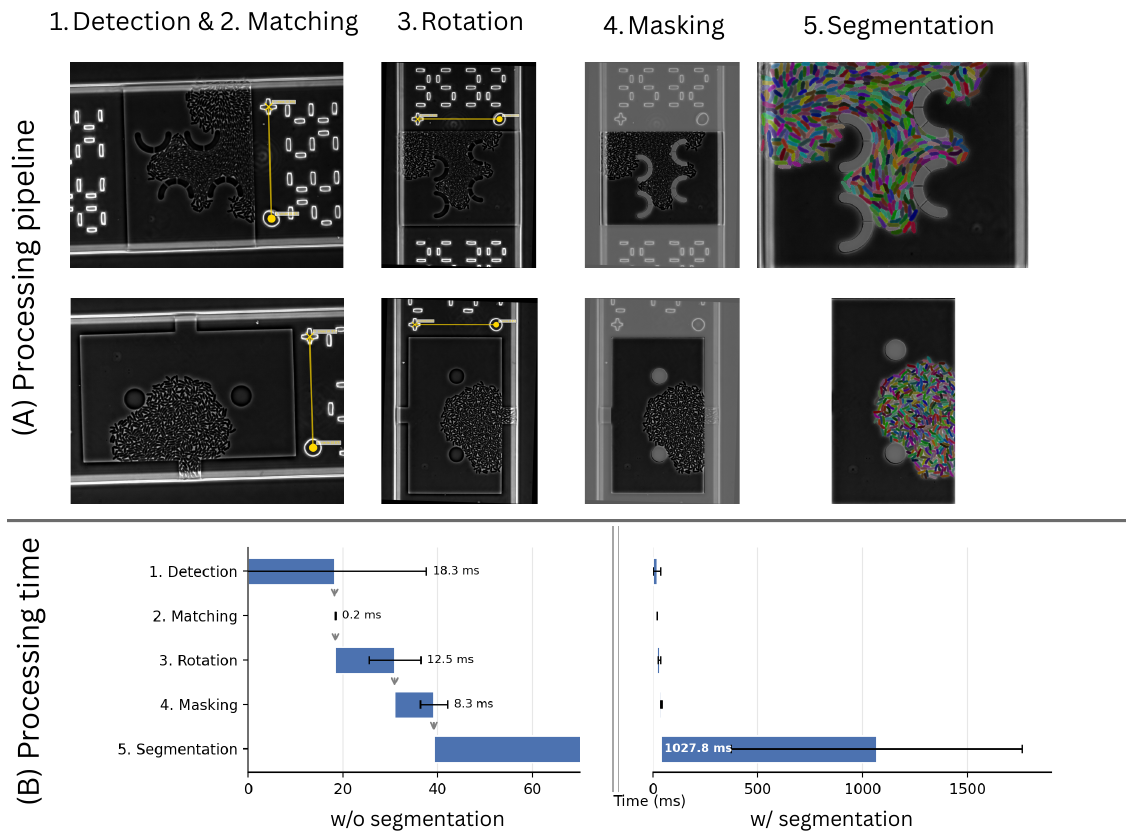}
    \caption{\textbf{\gls{dart}'s design-aware fine alignment and per-image processing pipeline.} (A)~Five-step image-processing workflow for two exemplary \gls{roi} designs comprising (1) fiducial marker detection, (2) marker-pair matching, (3) rotation and translation into the blueprint coordinate system, (4) blueprint-derived masking and cropping of microfluidic structures, and (5) cell segmentation. (B)~Mean processing time of each pipeline step measured during the live-cell validation experiment on the benchmarking hardware (see Methods). Steps 1--4 together required \SI[separate-uncertainty=true]{39.2(21.7)}{\milli\second} per image on average, whereas the segmentation step required \SI[separate-uncertainty=true]{1027.8(695.0)}{\milli\second} per image. Values are shown as mean $\pm$ one standard deviation. The detailed processing times are shown in \Cref{SItab:dart_masking_pipeline_details}.}  
    \label{fig:dart_masking_pipeline}
\end{figure}

While the coarse alignment places \glspl{roi} within the \gls{fov} for subsequent imaging, fine image-level alignment is performed separately during downstream image processing. \gls{dart}'s fine alignment registers the blueprint to each recorded microscopy image and thereby enables \gls{cad} design-aware masking of surrounding and interior microfluidic structures with sub-micron accuracy. To support real-time analysis, the per-image processing time must remain below the experimental acquisition interval. Given the \gls{roi} \gls{id} from the coarse alignment and the raw microscopy image, \gls{dart} performs five processing steps (\Cref{fig:dart_masking_pipeline}~A): (1)~detect fiducial markers, (2)~match valid marker pairs, (3)~compute the rotation and translation that align the image to the blueprint coordinate system, (4)~apply the blueprint-derived mask corresponding to the \gls{roi} \gls{id} to crop and exclude microfluidic structures, and (5)~segment cells in the preprocessed image using Cellpose-SAM~\cite{pachitariu_cellpose-sam_2025}. False-positive cell detections overlapping masked microfluidic structures are then removed automatically (see Methods).

Benchmarking this five-step per-image pipeline on the validation experiment and the benchmarking hardware (see Methods) yielded a mean execution time of \SI[separate-uncertainty=true]{39.2(21.7)}{\milli\second} per image without segmentation (\SI{25.5}{\fps}) and \SI[separate-uncertainty=true]{1.067(0.697)}{\second} per image including segmentation (\SI{0.94}{\fps}) (\Cref{fig:dart_masking_pipeline}~B and Methods). The first four processing steps contributed only a small and comparatively stable fraction of the runtime, whereas segmentation dominated the total processing time and showed greater variability. This variation arises because segmentation is applied to extracted \glspl{roi} of different sizes containing strongly varying cell numbers. In the validation experiment, the microscope acquired approximately one image every two seconds. Thus, the full pipeline, including segmentation, remained on average faster than the acquisition interval, enabling each image to be processed before the next one was recorded.

\gls{dart}'s fine alignment pipeline also incorporates automatic error detection. Specifically, the detected distance between each matched cross-circle marker pair is compared with the expected marker spacing from the blueprint. Marker pairs are declared valid when they are close to the expected marker spacing. Deviations beyond a configurable threshold flag likely marker detection failures, for example due to focus loss, marker deformation, or false-positive or false-negative marker detections. Automatic error detection automatically removes such images and prevents erroneous masks from propagating to downstream quantification. 

\subsection*{Live-cell experiment validation}

\begin{figure}
    \centering
    \includegraphics[width=0.9\linewidth]{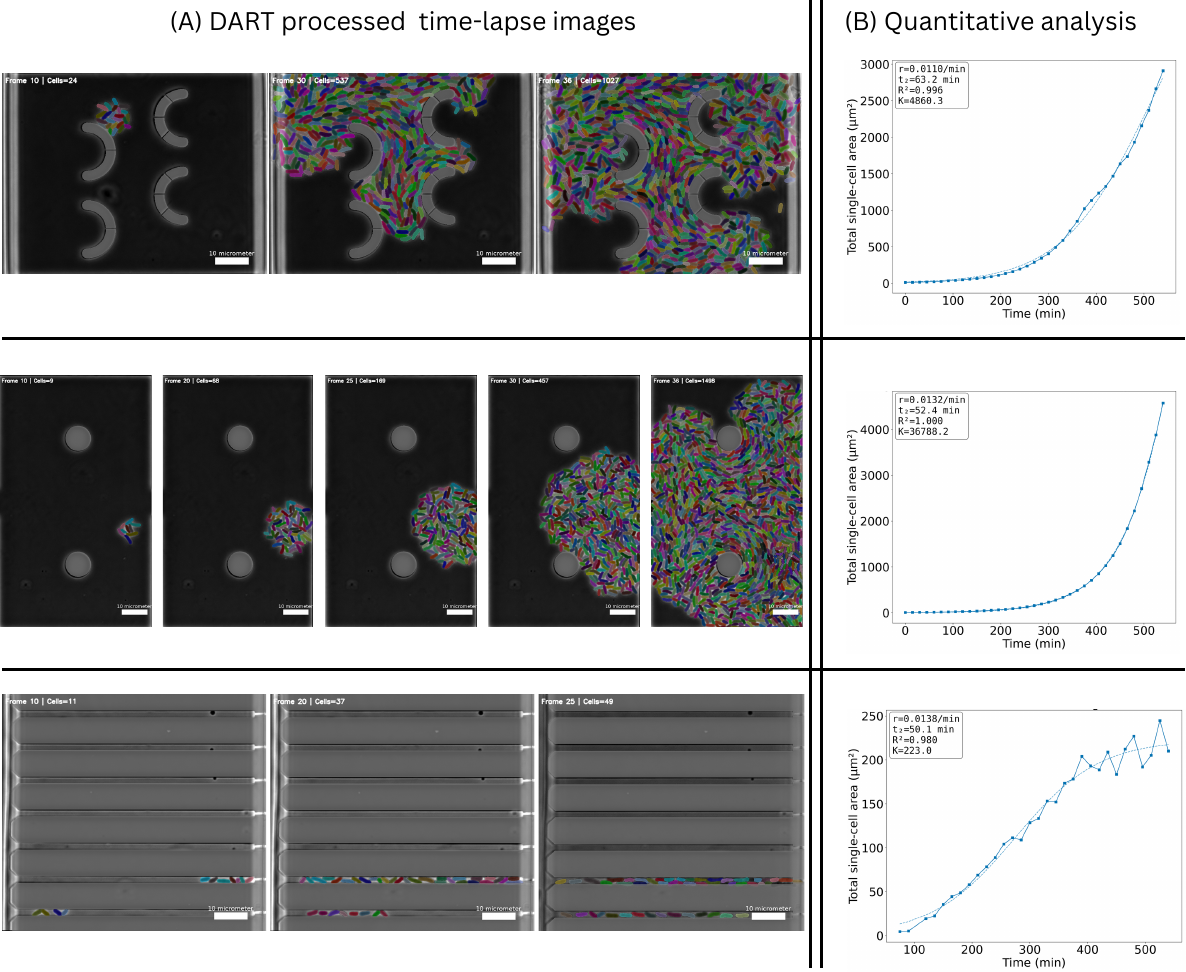}
    \caption{\textbf{Live-cell validation of the \gls{dart} pipeline on the \gls{sak} chip.} (A)~Representative time-lapse excerpts of \gls{cglut} populations cultivated in different \gls{sak} chamber geometries with complex interior structures. (B)~Temporal development of the \gls{tsca}, derived from automated \gls{dart} masking and cell segmentation (solid) and fitted logistic growth models (dashed) demonstrating automated extraction of quantitative growth dynamics.}  
     \label{fig:quantitative_MLCI_experiment}
\end{figure}

We validated the full \gls{dart} image-processing pipeline in a high-throughput live-cell experiment cultivating \gls{cglut} on the \gls{sak} chip (see Methods and Supplemental Movie S2). \Cref{fig:quantitative_MLCI_experiment}A shows representative time-lapse snapshots of developing cell populations in \gls{roi} designs containing the most complex interior structures. Microfluidic structures are highlighted in gray and segmented cells are shown as color-coded instances.  Detailed end-to-end processing throughput, cell counts, and real-time factor are provided in SI Text~\Cref{sec:SI_live_cell_experiment}. These examples demonstrate that \gls{dart} accurately masks surrounding and interior microfluidic features while preserving reliable cell segmentation in structurally complex cultivation geometries.

To assess whether the automated image analysis yields biologically interpretable outputs, we quantified the temporal development of the \gls{tsca} from the segmentation results (\Cref{fig:quantitative_MLCI_experiment}B). Fitting a logistic growth model (cf. Eq.~\eqref{eq:growth_model}) to these time series  yielded growth-rate estimates for the imaged cell populations. Thus, the live-cell experiment demonstrates end-to-end operation of the \gls{dart} pipeline on real cultivation data and shows that automated analysis on the \gls{sak} chip supports quantitative comparison of cell population dynamics across multiple \gls{roi} designs within a single experiment. Across the full live-cell experiment, \gls{dart} processed \num{1739} images in \SI{31}{\minute} and identified more than \num{500000} individual cells. Relative to the \SI{540}{\minute} acquisition time, this corresponds to a real-time factor of \num{17.4} (Supplemental Text~\ref{sec:SI_live_cell_experiment}). All images have been automatically processed using the full pipeline and while false-positive marker detections occurred, they have been successfully removed during the matching of valid marker pairs.

\section*{DISCUSSION}

\gls{dart} introduces a design-aware paradigm for microfluidic cultivation chips by linking the \gls{cad} blueprint to its physical implementation on the microscope stage. Its central advance is the recovery of structural design knowledge during experimentation, so that chip geometry becomes directly available to the image-analysis pipeline. This blueprint-to-device linkage enables fully automated, design-aware image processing and addresses two major bottlenecks in high-throughput live-cell imaging: throughput-scaling \gls{roi} localization during experiment setup and semi-automated post-hoc removal of microfluidic structures during image analysis. In the present study, \gls{dart} enabled sub-micron fiducial marker detection, throughput-independent \gls{roi} localization, and real-time-capable masking across diverse cultivation chamber geometries and chip layouts, as demonstrated on the \gls{sak} validation platform and in a live-cell experiment.

Coarse alignment replaces throughput-scaling manual \gls{roi} localization with a fixed-effort procedure requiring approximately five minutes of manual microscope operation. Because all \gls{roi} positions are inferred from the blueprint once the affine transformation is established, the workflow transfers across different \gls{roi} layouts and design classes without layout-specific reconfiguration. In practical terms, this reduces setup from a throughput-limiting manual search to a short calibration step, thereby preserving the operational advantage of high-density microfluidic chip designs. At the same time, localization accuracy remains limited at the stage level: a \num{90}th-percentile error corresponding to \SI{17.6}{\percent} of the \gls{fov} width may lead to partially imaged \glspl{roi} when the cultivation area occupies most of the \gls{fov}. For applications requiring tighter positioning, the stage map could be refined at the beginning of each experiment using additional reference measurements or corrected during acquisition using image-based position updates.

Fine alignment enables structure-aware masking of recorded images with sub-micron precision, including removal of interior microfluidic features in immediate proximity to cells. In runtime terms, fine alignment and masking contributed only a small fraction of the per-image processing time, whereas cell segmentation remained the dominant computational bottleneck. Thus, \gls{dart} already removes a major preprocessing barrier, and future runtime gains will likely depend primarily on faster segmentation models. Equally important, blueprint-driven masking eliminates the need for manual cropping or design-specific post-hoc exclusion of microfluidic structures, replacing a throughput-scaling preprocessing burden with a fully automated step. However, very narrow structures, such as mother-machine features with widths around \SI{1}{\micro\metre}, remain challenging because even small marker-localization errors may shift the projected mask sufficiently to occlude or expose parts of the cultivation area. In this regime, additional training data alone is unlikely to substantially improve alignment, as the observed \num{95}th-percentile marker-detection error of \SI{2.81}{\pixel} already approaches the precision of expert annotation at this image resolution. A more promising strategy may be to place additional markers around the \gls{roi}, which could further improve local alignment and robustness against defective or weakly visible markers. Beyond alignment itself, the built-in error detection could also support quality control of fabricated chips by identifying problematic regions before the experiment.

The live-cell validation further shows that \gls{dart} is not only a design concept, but an end-to-end hardware--software workflow that operates on real cultivation data and yields biologically interpretable outputs. By combining a marker-encoded chip design with blueprint-driven image analysis, \gls{dart} makes a broad range of \gls{roi} architectures more experimentally practical, because structural complexity no longer requires design-specific post-hoc preprocessing or retraining of image-analysis models. Across the major chamber geometries represented on the \gls{sak}, heterogeneous chip designs were therefore analyzed within one common workflow rather than requiring geometry-specific analysis pipelines. In this sense, \gls{dart} contributes both a device-level design principle and a directly usable analysis workflow. The use of established fabrication routes and open-source software further lowers the barrier to adoption in other \gls{mlci} laboratories.

At the same time, \gls{dart} establishes a practical foundation for more autonomous microscopy workflows. By combining structural awareness with real-time capable image analysis, \gls{dart} provides key prerequisites for adaptive experimentation, including event-triggered~\cite{chiron_cybersco_2022}, outcome-driven~\cite{chiron_cybersco_2022,passmore_closedloop_2025}, and closed-loop experiments~\cite{lugagne_deep_2024}. By using blueprint information during image analysis, \gls{dart} brings interpretation substantially closer to the timescale of data acquisition and thus provides an important enabling step toward smart microscopy in structured microfluidic systems.

More broadly, \gls{dart} demonstrates that structural knowledge encoded during chip design is recoverable at runtime and directly available for automated real-time image analysis. Although we applied this concept here to microbial microfluidic cultivation chips, the same principle may extend to other structured live-cell imaging platforms in which the spatial layout is digitally specified before experimentation. In this sense, \gls{dart} provides not only a practical solution for high-throughput microfluidic live-cell imaging, but also an example of how design-aware analysis helps bringing image interpretation closer to the timescale of data acquisition. Such real-time capable analysis is an important prerequisite for future event-based, closed-loop, and outcome-driven microscopy experiments.

\section*{METHODS}

\subsection*{Converting a microfluidic design into a DART chip}
\label{sec:M_dart_conversion}

The \gls{dart} paradigm extends existing microfluidic \gls{cad} designs by adding fiducial markers and numeric \glspl{id}. We parse blueprint designs provided in the \gls{cif} file format, which encodes the chip layout as a hierarchical tree of symbols composed of geometric primitives such as boxes, polygons, and circles. We then identify the symbols corresponding to \glspl{roi} and augment each of them with a cross marker, a circle marker, and a unique \gls{id} number placed at configurable positions and distances. The augmented design is written back in \gls{cif} format and used for chip fabrication.

For the fiducial markers, we selected a cross-and-circle design with a diameter of \SI{8}{\micro\metre}. The cross bars were designed with a width of \SI{2}{\micro\metre}. These dimensions were chosen to ensure reliable fabrication, as structures below \SI{1}{\micro\metre} proved difficult to manufacture reproducibly using standard microfabrication processes.

\subsection*{Microfluidic chip design and fabrication}
\label{sec:M_chip_fabrication}

The \gls{sak} chip was designed in CleWin5 (WieWeb Software, Hengelo, The Netherlands) and converted into a \gls{dart} design using the procedure described above. The resulting layout was used to fabricate a \SI{100}{\milli\metre} silicon wafer mold. The \SI{1}{\micro\metre}-deep cultivation structures were patterned in SU-8 photoresist by electron-beam lithography (Raith EBPG~5200), whereas the larger supply-channel structures were fabricated using a direct-write laser system (DWL~66+, Heidelberg Instruments).

\gls{pdms} chips were cast from the silicon mold using a \num{10}:\num{1} mixture of base and curing agent, degassed at \SI{200}{\milli\bar} for \SI{45}{\minute}, and cured at \SI{80}{\degreeCelsius} for \SI{2}{\hour}. 
The cured chips were cut along printed cutting lines, treated with pentane and acetone for \SI{90}{\minute} to improve durability~\cite{Gruenberger2013}, and punched with inlet and outlet holes (\SI{0.5}{\milli\metre} diameter; World Precision Instruments). Finally, the chips were bonded to glass substrates (D263\textsuperscript{\textregistered}Eco, \num{39.5}~$\times$~\num{34.5}~$\times$~\SI{0.175}{\milli\metre}, Schott AG) using oxygen plasma (Femto Plasma Cleaner, Diener Electronics).

\subsection*{Microfluidic live-cell imaging}
\label{sec:M_experiment}

\gls{cglut} (ATCC~13032) was used for experimental validation of the \gls{dart} pipeline. Cells were cultivated in Brain Heart Infusion (BHI) medium at \SI{30}{\degreeCelsius} and \SI{120}{\rpm} in \SI{100}{\milli\litre} baffled shake flasks containing \SI{20}{\milli\litre} culture volume. Pre-cultures were grown for \SI{16}{\hour}, and main cultures were inoculated at an optical density of \num{0.2}. After \SI{3}{\hour} of exponential growth, cells were harvested for chip inoculation to minimize lag-phase effects during on-chip cultivation. The optical density of the inoculation suspension was adjusted between \num{0.1} and \num{1.0}, depending on the desired initial cell density.

Time-lapse imaging was performed on a Nikon Ti inverted microscope platform using NIS-Elements software. Phase-contrast images were acquired at \SI{100}~$\times$ magnification using two cameras: an Andor sCMOS Neo 5.5 camera (native resolution \SI{2560}~$\times$~\SI{2160}{\pixel}; pixel resolution: \SI{0.066}{\micro\metre\per\pixel}) and an Andor DR-328G-C02-SIL camera (native resolution \num{1392}~$\times$~\SI{1040}{\pixel}; pixel size \SI{0.093}{\micro\metre\per\pixel}). The Neo 5.5 camera was used to record the coarse alignment dataset, whereas the DR-328G-C02-SIL camera was used for the live-cell validation experiment.

\subsection*{Marker detection data annotation and composition}
\label{sec:M_yolo_data}

For training the \gls{yolo}-based marker detector, we recorded phase-contrast images of \glspl{roi} on the \gls{sak} chip using a \num{100}~$\times$ objective (native image resolution \num{2560}~$\times$~\SI{2160}{\pixel}; \SI{0.066}{\micro\metre\per\pixel}).

From these recordings, a random subset of \num{319} images was selected for manual marker annotation. Annotation was performed in Label Studio (\url{https://labelstud.io/}) using \gls{sam}-assisted interactive segmentation~\cite{ravi_sam2_2024}. Specifically, point annotations were placed manually on fiducial markers in each image, after which \gls{sam} predicted the corresponding marker masks; these masks were manually corrected where necessary. Defective or partially visible marker instances were not annotated, and out-of-focus images were excluded. In total, \num{472} cross markers and \num{470} circle markers were annotated. The final dataset was split into training, validation, and test sets (\SI{60}{\percent}, \SI{15}{\percent}, and \SI{25}{\percent}, respectively). Detailed dataset statistics are provided in \Cref{SItab:dataset-statistics}, and the annotated dataset has been made publicly available.

\subsection*{YOLO training configuration}
\label{sec:M_yolo_config}

\gls{yolo} models were trained using the Ultralytics framework~\cite{jocher_ultralytics_2023} with a batch size of \num{16}. Each model configuration was trained for \num{100} epochs. Training was performed on a workstation equipped with two AMD EPYC 7752 \num{64}-core CPUs, an NVIDIA A100 GPU with \SI{40}{\giga\byte} memory, and \SI{512}{\giga\byte} RAM. Training time per model configuration was approximately \SI{26}{\minute}.

Built-in \gls{yolo} augmentations were combined with custom microscopy-specific augmentations implemented using the \texttt{albumentations} library to simulate suboptimal imaging conditions, including defocus, noise, and illumination variation. The full augmentation configuration and example training batches are provided in \Cref{SItab:augmentations} and \Cref{SIfig:train-batch}.

\subsection*{Coarse alignment validation}
\label{sec:M_map_validation}

After determining the affine transformation from three calibration images, the microscope stage was moved to each \gls{roi} position predicted by the map. At each position, a phase-contrast image was recorded and the corresponding stage coordinates were stored (Neo 5.5 camera; pixel size \SI{0.066}{\micro\metre\per\pixel}). Marker detection and marker-pair matching were then applied to each image. The image-derived \gls{roi} center was reconstructed from the detected cross-marker position by adding the known blueprint offset between the cross marker and the \gls{roi} centroid, rotated according to the detected marker orientation. This center position was then converted from image coordinates into stage coordinates using the camera pixel size and the recorded stage position.

The reconstructed image-derived \gls{roi} center was compared with the map-predicted \gls{roi} center for the same \gls{roi} \gls{id}. Because the coarse alignment maps blueprint and stage coordinates into the same reference frame, both positions are directly comparable. The position error for each \gls{roi} was defined as the Euclidean distance between the image-derived and map-predicted center positions in micrometers. The measured alignment error reflects contributions from microscope stage positioning inaccuracy, fabrication tolerances of the \gls{pdms} chip, and local deformation of the soft \gls{pdms} material relative to the \gls{cad} blueprint. We report the median and \num{90}th-percentile absolute position error across all \glspl{roi} with a valid detected marker pair; images without a valid marker pair were excluded.

\subsection*{Fine alignment pipeline}
\label{sec:M_masking}

The fine alignment pipeline registers the microfluidic blueprint to each recorded microscopy image using the fiducial markers and thereby enables automated, design-aware image processing. The pipeline comprises the five steps outlined in \Cref{fig:dart_masking_pipeline}A.

First, fiducial marker detection is performed. The selected \gls{yolo}v26 model (detection task, size~s, input resolution \SI{1280}{\pixel}) is applied to the full microscopy image at a confidence threshold of \num{0.5}, yielding bounding-box center coordinates for detected cross and circle markers.

Second, marker-pair matching and rotation angle estimation are performed. All detected cross--circle combinations are evaluated as candidate pairs and scored by the absolute difference between the detected marker distance and the blueprint-expected marker distance. Candidate pairs deviating by more than \SI{60}{\pixel} are discarded, and the remaining pairs are matched greedily in ascending distance order without reusing markers. If multiple valid pairs are found, their implied rotation angles are computed; pairs deviating by more than \SI{5}{\degree} from the mean angle are discarded, and the mean angle of the remaining consistent pairs is used for image rotation.

Third, the image is rotated into the blueprint coordinate system. The rotation angle is computed from the detected and expected cross-to-circle unit vectors, and the full image is rotated around its center using GPU-accelerated transformations from \texttt{Kornia}~\cite{riba_kornia_2020}. If GPU-based processing is unavailable, OpenCV~\cite{bradski_opencv_2000} \texttt{warpAffine} with bilinear interpolation is used as a fallback. Marker coordinates are updated accordingly after rotation.

Fourth, design-aware masking and cropping are performed. The polygon corresponding to the detected \gls{roi} \gls{id} is pre-loaded from the chip configuration derived from the \gls{cif} design. This polygon is translated to the detected cross-marker position and rasterized into a binary mask using \texttt{Shapely}~\cite{gillies_shapely_2007} and \texttt{Rasterio}~\cite{gillies_rasterio_2013}. The mask is applied to the rotated image, and both image and mask are cropped to the bounding box of the \gls{roi} polygon. The resulting masks for all eight \gls{roi} designs are shown in \Cref{SIfig:dmc_rois_masked}.

Fifth, cell segmentation is performed on the cropped and masked single-channel phase-contrast \gls{roi} image using Cellpose-SAM~\cite{pachitariu_cellpose-sam_2025} with default parameters. Restricting segmentation to the cropped \gls{roi}, rather than the full microscopy image, reduces computational load approximately in proportion to the \gls{roi}-to-image area ratio.

\subsection*{Pipeline benchmarking and timings}
\label{sec:M_benchmarking}

Each pipeline step (marker detection, marker matching, image rotation, masking and cropping, and cell segmentation) was timed individually using \texttt{time.perf\_counter()} to record wall-clock time. Timings were collected during processing of the full live-cell experiment (\num{1739} images) using \path{process_folder.py}; no separate benchmarking run was performed. Images were processed sequentially with batch size~\num{1} to reflect the intended real-time, single-image processing scenario.

All pipeline benchmarks were performed on the training server. We report the mean and standard deviation of the per-step wall-clock time across all successfully processed images. Timing results grouped by chamber geometry type, together with the detailed per-chamber breakdown, are provided in \Cref{SItab:dart_masking_pipeline_details}.

\subsection*{Growth analysis and total segmented cell area}
\label{sec:M_growth_analysis}

For each recorded image, the total segmented cell area (\gls{tsca}) was defined as the sum of the areas of all individually segmented cells. Cell areas were derived from the Cellpose-SAM instance segmentation masks by counting the number of segmented cell pixels and converting this value into physical area using the camera-specific pixel calibration. Thus, the segmented area was expressed in \SI{}{\micro\metre\squared}. For each cultivation chamber and time point, the areas of all segmented cells were summed to obtain the \gls{tsca} time series.

A logistic growth model~\cite{zwietering_modeling_1990}
\begin{equation}
    N(t) = \frac{K}{1 + \frac{K - N_0}{N_0} \cdot \exp(-r \cdot t)}
    \label{eq:growth_model}
\end{equation}
was fitted to each \gls{tsca} time series, where $N_0$ denotes the initial \gls{tsca}, $r$ the intrinsic growth rate, and $K$ the carrying capacity corresponding to the maximum \gls{tsca} at chamber saturation. Model parameters were estimated by nonlinear least-squares fitting using \path{scipy.optimize.curve_fit}~\cite{virtanen_scipy_2020}. Initial parameter values were selected as follows: $N_0$ was set to the mean of the first quarter of time points, $K$ to $1.1 \times \max(N)$, and $r$ to the log-linear slope of the first half of the time series. Parameter values were restricted to $(0, \infty)$. We report the key performance indicators for microbial growth: the intrinsic growth rate $r$, the doubling time $t_2 = \ln(2)/r$, the carrying capacity $K$, and the coefficient of determination $R^2$ computed on the original \gls{tsca} scale.

\section*{RESOURCE AVAILABILITY}

\subsection*{Lead contact}

Requests for further information and resources should be directed to and will be fulfilled by the lead contact, Katharina Nöh (\href{mailto:k.noeh@fz-juelich.de}{k.noeh@fz-juelich.de}).

\subsection*{Materials availability}

No special or new materials have been used or developed in this work.

\subsection*{Data and code availability}

\begin{itemize}
    \item The annotated live-cell imaging dataset for training the \gls{yolo}-based marker detector is available at \url{https://fz-juelich.sciebo.de/s/cw7jMk3ZWzB2JbF}.
    \item The coarse alignment validation dataset is available at \url{https://fz-juelich.sciebo.de/s/9bRnA364D8KgixE}.
    \item The live-cell imaging dataset is available at\\ \url{https://fz-juelich.sciebo.de/s/Tq5SW76WG9zqMJi} and also contains the \gls{sak} chip \gls{cad} design file in \gls{cif} format.
    \item The \gls{dart} software for coarse and fine alignment is available at \url{https://github.com/SMLCI/DART-MLCI} and on PyPI.
\end{itemize}

\section*{ACKNOWLEDGMENTS}

This work was supported by the President's Initiative and Networking Funds of the Helmholtz Association of German Research Centres via grant [EMSIG ZT-I-PF-04-44], the Bioeconomy Science Center via grant [HAIPSs3D PhD-MP\_2024\_05], supported by the Ministry of Culture and Science of the State of North Rhine-Westphalia, and received funding from the Helmholtz Association of German Research Centres within the Helmholtz School for Data Science in Life, Earth, and Energy (HDS-LEE).

\section*{AUTHOR CONTRIBUTIONS}

Conceptualization, J.S.; Methodology, J.S. and L.S.; Software, J.S. and L.S.; Validation, J.S. and M.P.; Formal analysis, J.S.; Investigation, J.S., L.S., and M.P.; Resources, D.K.; Data curation, J.S., L.S., and M.P.; Writing – original draft, J.S., and L.S.; Writing – review \& editing, J.S., K.N., M.P., D.K., and H.S.;
Visualization, J.S.; Supervision, K.N.; Project administration, K.N.; Funding acquisition, K.N., H.S., D.K., and J.S.

\section*{DECLARATION OF INTERESTS}

The authors declare no competing interests.

\section*{DECLARATION OF GENERATIVE AI AND AI-ASSISTED TECHNOLOGIES}

During the preparation of this work, the authors used Claude in order to improve readability. After using this tool, the authors reviewed and edited the content as needed and take full responsibility for the content of the publication.

\section*{SUPPLEMENTAL INFORMATION INDEX}

\begin{description}
  \item Document S1. 
    Supplemental methods consisting of Text S1~-~S3, Figures S1~-S4, and Tables S1~-~S5 (PDF).
  \item Movie S1. 
    Detailed \gls{dart} image-processing workflow across the eight \gls{sak} \gls{roi} designs, showing marker detection, blueprint alignment, masking, and segmentation overlay.
  \item Movie S2. 
    Real-time \gls{dart} image-processing workflow applied to time-lapse recordings from the \gls{cglut} cultivation experiment, showing segmented cells across \gls{sak} \glspl{roi}.
\end{description}

\newpage

\bibliography{references}

\begin{thebibliography}{38}
\providecommand{\natexlab}[1]{#1}
\providecommand{\url}[1]{\texttt{#1}}
\providecommand{\href}[2]{#2}
\providecommand{\path}[1]{#1}
\providecommand{\DOIprefix}{doi: }
\providecommand{\ArXivprefix}{arXiv: }
\providecommand{\URLprefix}{URL: }
\providecommand{\Pubmedprefix}{pmid: }
\providecommand{\doi}[1]{\href{http://dx.doi.org/#1}{\path{#1}}}
\providecommand{\Pubmed}[1]{\href{pmid:#1}{\path{#1}}}
\providecommand{\BIBand}{and}
\providecommand{\bibinfo}[2]{#2}
\ifx\xfnm\undefined \def\xfnm[#1]{\unskip,\space#1}\fi
\makeatletter\def\@biblabel#1{#1.}\makeatother
\bibitem[{Wang et~al.(2010)Wang, Robert, Pelletier, Dang, Taddei, Wright and Jun}]{wang_robust_2010}
\bibinfo{author}{Wang, P.}, \bibinfo{author}{Robert, L.}, \bibinfo{author}{Pelletier, J.}, \bibinfo{author}{Dang, W.L.}, \bibinfo{author}{Taddei, F.}, \bibinfo{author}{Wright, A.}, and \bibinfo{author}{Jun, S.} (\bibinfo{year}{2010}). \bibinfo{title}{Robust {Growth} of {Escherichia} coli}.
\newblock \bibinfo{journal}{Current Biology} \emph{\bibinfo{volume}{20}}, \bibinfo{pages}{1099--1103}. \DOIprefix\doi{10.1016/j.cub.2010.04.045}.
\bibitem[{Grünberger et~al.(2012)Grünberger, Paczia, Probst, Schendzielorz, Eggeling, Noack, Wiechert and Kohlheyer}]{grunberger_disposable_2012}
\bibinfo{author}{Grünberger, A.}, \bibinfo{author}{Paczia, N.}, \bibinfo{author}{Probst, C.}, \bibinfo{author}{Schendzielorz, G.}, \bibinfo{author}{Eggeling, L.}, \bibinfo{author}{Noack, S.}, \bibinfo{author}{Wiechert, W.}, and \bibinfo{author}{Kohlheyer, D.} (\bibinfo{year}{2012}). \bibinfo{title}{A disposable picolitre bioreactor for cultivation and investigation of industrially relevant bacteria on the single cell level}.
\newblock \bibinfo{journal}{Lab on a Chip} \emph{\bibinfo{volume}{12}}, \bibinfo{pages}{2060}. \DOIprefix\doi{10.1039/c2lc40156h}.
\bibitem[{Gr\"unberger et~al.(2013)Gr\"unberger, Probst, Heyer, Wiechert, Frunzke and Kohlheyer}]{grunberger_microfluidic_2013}
\bibinfo{author}{Gr\"unberger, A.}, \bibinfo{author}{Probst, C.}, \bibinfo{author}{Heyer, A.}, \bibinfo{author}{Wiechert, W.}, \bibinfo{author}{Frunzke, J.}, and \bibinfo{author}{Kohlheyer, D.} (\bibinfo{year}{2013}). \bibinfo{title}{Microfluidic picoliter bioreactor for microbial single-cell analysis: Fabrication, system setup, and operation}.
\newblock \bibinfo{journal}{Journal of Visualized Experiments} pp. \bibinfo{pages}{e50560}. \DOIprefix\doi{10.3791/50560}.
\bibitem[{Kasahara et~al.(2025{\natexlab{a}})Kasahara, Seiffarth, Stute, von Lieres, Drepper, N{\"{o}}h and Kohlheyer}]{Kasahara2025}
\bibinfo{author}{Kasahara, K.}, \bibinfo{author}{Seiffarth, J.}, \bibinfo{author}{Stute, B.}, \bibinfo{author}{von Lieres, E.}, \bibinfo{author}{Drepper, T.}, \bibinfo{author}{N{\"{o}}h, K.}, and \bibinfo{author}{Kohlheyer, D.} (\bibinfo{year}{2025}{\natexlab{a}}). \bibinfo{title}{{Unveiling microbial single-cell growth dynamics under rapid periodic oxygen oscillations}}.
\newblock \bibinfo{journal}{Lab on a Chip} \emph{\bibinfo{volume}{25}}, \bibinfo{pages}{2234--2246}. \DOIprefix\doi{10.1039/D5LC00065C}.
\bibitem[{Kaiser et~al.(2018)Kaiser, Jug, Julou, Deshpande, Pfohl, Silander, Myers and van Nimwegen}]{kaiser_monitoring_2018}
\bibinfo{author}{Kaiser, M.}, \bibinfo{author}{Jug, F.}, \bibinfo{author}{Julou, T.}, \bibinfo{author}{Deshpande, S.}, \bibinfo{author}{Pfohl, T.}, \bibinfo{author}{Silander, O.K.}, \bibinfo{author}{Myers, G.}, and \bibinfo{author}{van Nimwegen, E.} (\bibinfo{year}{2018}). \bibinfo{title}{Monitoring single-cell gene regulation under dynamically controllable conditions with integrated microfluidics and software}.
\newblock \bibinfo{journal}{Nature Communications} \emph{\bibinfo{volume}{9}}, \bibinfo{pages}{212}. \DOIprefix\doi{10.1038/s41467-017-02505-0}.
\bibitem[{Grünberger et~al.(2015)Grünberger, Probst, Helfrich, Nanda, Stute, Wiechert, von Lieres, Nöh, Frunzke and Kohlheyer}]{grunberger_spatiotemporal_2015}
\bibinfo{author}{Grünberger, A.}, \bibinfo{author}{Probst, C.}, \bibinfo{author}{Helfrich, S.}, \bibinfo{author}{Nanda, A.}, \bibinfo{author}{Stute, B.}, \bibinfo{author}{Wiechert, W.}, \bibinfo{author}{von Lieres, E.}, \bibinfo{author}{Nöh, K.}, \bibinfo{author}{Frunzke, J.}, and \bibinfo{author}{Kohlheyer, D.} (\bibinfo{year}{2015}). \bibinfo{title}{Spatiotemporal microbial single-cell analysis using a high-throughput microfluidics cultivation platform}.
\newblock \bibinfo{journal}{Cytometry Part A} \emph{\bibinfo{volume}{87}}, \bibinfo{pages}{1101--1115}. \DOIprefix\doi{10.1002/cyto.a.22779}.
\bibitem[{Helfrich et~al.(2015)Helfrich, Pfeifer, Krämer, Sachs, Wiechert, Kohlheyer, Nöh and Frunzke}]{helfrich_live_2015}
\bibinfo{author}{Helfrich, S.}, \bibinfo{author}{Pfeifer, E.}, \bibinfo{author}{Krämer, C.}, \bibinfo{author}{Sachs, C.C.}, \bibinfo{author}{Wiechert, W.}, \bibinfo{author}{Kohlheyer, D.}, \bibinfo{author}{Nöh, K.}, and \bibinfo{author}{Frunzke, J.} (\bibinfo{year}{2015}). \bibinfo{title}{Live cell imaging of {SOS} and prophage dynamics in isogenic bacterial populations}.
\newblock \bibinfo{journal}{Molecular Microbiology} \emph{\bibinfo{volume}{98}}, \bibinfo{pages}{636--650}. \DOIprefix\doi{10.1111/mmi.13147}.
\bibitem[{Sachs et~al.(2016)Sachs, Gr\"unberger, Helfrich, Probst, Wiechert, Kohlheyer and N\"oh}]{sachs_image_2016}
\bibinfo{author}{Sachs, C.C.}, \bibinfo{author}{Gr\"unberger, A.}, \bibinfo{author}{Helfrich, S.}, \bibinfo{author}{Probst, C.}, \bibinfo{author}{Wiechert, W.}, \bibinfo{author}{Kohlheyer, D.}, and \bibinfo{author}{N\"oh, K.} (\bibinfo{year}{2016}). \bibinfo{title}{Image-based single cell profiling: High-throughput processing of mother machine experiments}.
\newblock \bibinfo{journal}{PLOS ONE} \emph{\bibinfo{volume}{11}}, \bibinfo{pages}{e0163453}. \DOIprefix\doi{10.1371/journal.pone.0163453}.
\bibitem[{Blöbaum et~al.(2023)Blöbaum, Täuber and Grünberger}]{blobaum_protocol_2023}
\bibinfo{author}{Blöbaum, L.}, \bibinfo{author}{Täuber, S.}, and \bibinfo{author}{Grünberger, A.} (\bibinfo{year}{2023}). \bibinfo{title}{Protocol to perform dynamic microfluidic single-cell cultivation of {C}. glutamicum}.
\newblock \bibinfo{journal}{STAR Protocols} \emph{\bibinfo{volume}{4}}, \bibinfo{pages}{102436}. \DOIprefix\doi{10.1016/j.xpro.2023.102436}.
\bibitem[{O'Connor et~al.(2022)O'Connor, Alnahhas, Lugagne and Dunlop}]{oconnor_delta_2022}
\bibinfo{author}{O'Connor, O.M.}, \bibinfo{author}{Alnahhas, R.N.}, \bibinfo{author}{Lugagne, J.B.}, and \bibinfo{author}{Dunlop, M.J.} (\bibinfo{year}{2022}). \bibinfo{title}{{DeLTA} 2.0: {A} deep learning pipeline for quantifying single-cell spatial and temporal dynamics}.
\newblock \bibinfo{journal}{PLOS Computational Biology} \emph{\bibinfo{volume}{18}}, \bibinfo{pages}{e1009797}. \DOIprefix\doi{10.1371/journal.pcbi.1009797}.
\bibitem[{Prangemeier et~al.(2022)Prangemeier, Wildner, Fran\c{c}ani, Reich and Koeppl}]{prangemeier_yeast_2022}
\bibinfo{author}{Prangemeier, T.}, \bibinfo{author}{Wildner, C.}, \bibinfo{author}{Fran\c{c}ani, A.O.}, \bibinfo{author}{Reich, C.}, and \bibinfo{author}{Koeppl, H.} (\bibinfo{year}{2022}). \bibinfo{title}{Yeast cell segmentation in microstructured environments with deep learning}.
\newblock \bibinfo{journal}{Biosystems} \emph{\bibinfo{volume}{211}}, \bibinfo{pages}{104557}. \DOIprefix\doi{10.1016/j.biosystems.2021.104557}.
\bibitem[{Zhou et~al.(2023)Zhou, Chen, Fu and Yan}]{zhou_computer_2023}
\bibinfo{author}{Zhou, S.}, \bibinfo{author}{Chen, B.}, \bibinfo{author}{Fu, E.S.}, and \bibinfo{author}{Yan, H.} (\bibinfo{year}{2023}). \bibinfo{title}{Computer vision meets microfluidics: a label-free method for high-throughput cell analysis}.
\newblock \bibinfo{journal}{Microsystems \& Nanoengineering} \emph{\bibinfo{volume}{9}}. \DOIprefix\doi{10.1038/s41378-023-00562-8}.
\bibitem[{Cutler et~al.(2022)Cutler, Stringer, Lo, Rappez, Stroustrup, Brook~Peterson, Wiggins and Mougous}]{cutler_omnipose_2022}
\bibinfo{author}{Cutler, K.J.}, \bibinfo{author}{Stringer, C.}, \bibinfo{author}{Lo, T.W.}, \bibinfo{author}{Rappez, L.}, \bibinfo{author}{Stroustrup, N.}, \bibinfo{author}{Brook~Peterson, S.}, \bibinfo{author}{Wiggins, P.A.}, and \bibinfo{author}{Mougous, J.D.} (\bibinfo{year}{2022}). \bibinfo{title}{Omnipose: a high-precision morphology-independent solution for bacterial cell segmentation}.
\newblock \bibinfo{journal}{Nature Methods} pp. \bibinfo{pages}{1--11}. \DOIprefix\doi{10.1038/s41592-022-01639-4}.
\bibitem[{Stringer et~al.(2020)Stringer, Wang, Michaelos and Pachitariu}]{stringer_cellpose_2021}
\bibinfo{author}{Stringer, C.}, \bibinfo{author}{Wang, T.}, \bibinfo{author}{Michaelos, M.}, and \bibinfo{author}{Pachitariu, M.} (\bibinfo{year}{2020}). \bibinfo{title}{Cellpose: a generalist algorithm for cellular segmentation}.
\newblock \bibinfo{journal}{Nature Methods} \emph{\bibinfo{volume}{18}}, \bibinfo{pages}{100--106}. \DOIprefix\doi{10.1038/s41592-020-01018-x}.
\bibitem[{Pachitariu et~al.(2025)Pachitariu, Rariden and Stringer}]{pachitariu_cellpose-sam_2025}
\bibinfo{author}{Pachitariu, M.}, \bibinfo{author}{Rariden, M.}, and \bibinfo{author}{Stringer, C.} (\bibinfo{year}{2025}). \bibinfo{title}{Cellpose-{SAM}: superhuman generalization for cellular segmentation}.
\newblock \bibinfo{journal}{bioRxiv}. \DOIprefix\doi{10.1101/2025.04.28.651001}.
\bibitem[{Marks et~al.(2025)Marks, Israel, Van~Valen et~al.}]{marks_cellsam_2025}
\bibinfo{author}{Marks, U.}, \bibinfo{author}{Israel, R.}, \bibinfo{author}{Van~Valen, D.} et~al. (\bibinfo{year}{2025}). \bibinfo{title}{A foundation model for cellular segmentation}.
\newblock \bibinfo{journal}{Nature Methods}. \DOIprefix\doi{10.1038/s41592-025-02879-w}.
\bibitem[{Archit et~al.(2025)Archit, Freckmann, Nair, Khalid, Hilt, Rajashekar, Freitag, Teuber, Buckley, von Haaren, Gupta, Dengel, Ahmed and Pape}]{archit_microsam_2024}
\bibinfo{author}{Archit, A.}, \bibinfo{author}{Freckmann, L.}, \bibinfo{author}{Nair, S.}, \bibinfo{author}{Khalid, N.}, \bibinfo{author}{Hilt, P.}, \bibinfo{author}{Rajashekar, V.}, \bibinfo{author}{Freitag, M.}, \bibinfo{author}{Teuber, C.}, \bibinfo{author}{Buckley, G.}, \bibinfo{author}{von Haaren, S.}, \bibinfo{author}{Gupta, S.}, \bibinfo{author}{Dengel, A.}, \bibinfo{author}{Ahmed, S.}, and \bibinfo{author}{Pape, C.} (\bibinfo{year}{2025}). \bibinfo{title}{Segment anything for microscopy}.
\newblock \bibinfo{journal}{Nature Methods}. \DOIprefix\doi{10.1038/s41592-024-02580-4}.
\bibitem[{Kasahara et~al.(2025{\natexlab{b}})Kasahara, Seiffarth, Stute, von Lieres, Drepper, Nöh and Kohlheyer}]{kasahara_unveiling_2025}
\bibinfo{author}{Kasahara, K.}, \bibinfo{author}{Seiffarth, J.}, \bibinfo{author}{Stute, B.}, \bibinfo{author}{von Lieres, E.}, \bibinfo{author}{Drepper, T.}, \bibinfo{author}{Nöh, K.}, and \bibinfo{author}{Kohlheyer, D.} (\bibinfo{year}{2025}{\natexlab{b}}). \bibinfo{title}{Unveiling microbial single-cell growth dynamics under rapid periodic oxygen oscillations}.
\newblock \bibinfo{journal}{Lab on a Chip} \emph{\bibinfo{volume}{25}}, \bibinfo{pages}{2234--2246}. \DOIprefix\doi{10.1039/D5LC00065C}.
\bibitem[{Witting et~al.(2025)Witting, Seiffarth, Stute, Schulze, Hofer, Nöh, Eisenhut, Weber, von Lieres and Kohlheyer}]{witting_microfluidic_2025}
\bibinfo{author}{Witting, L.}, \bibinfo{author}{Seiffarth, J.}, \bibinfo{author}{Stute, B.}, \bibinfo{author}{Schulze, T.}, \bibinfo{author}{Hofer, J.M.}, \bibinfo{author}{Nöh, K.}, \bibinfo{author}{Eisenhut, M.}, \bibinfo{author}{Weber, A.P.M.}, \bibinfo{author}{von Lieres, E.}, and \bibinfo{author}{Kohlheyer, D.} (\bibinfo{year}{2025}). \bibinfo{title}{A microfluidic system for the cultivation of cyanobacteria with precise light intensity and {CO2} control: enabling growth data acquisition at single-cell resolution}.
\newblock \bibinfo{journal}{Lab on a Chip} \emph{\bibinfo{volume}{25}}, \bibinfo{pages}{319--329}. \DOIprefix\doi{10.1039/D4LC00567H}.
\bibitem[{Dal~Co et~al.(2020)Dal~Co, van Vliet, Kiviet, Schlegel and Ackermann}]{dal_co_short-range_2020}
\bibinfo{author}{Dal~Co, A.}, \bibinfo{author}{van Vliet, S.}, \bibinfo{author}{Kiviet, D.J.}, \bibinfo{author}{Schlegel, S.}, and \bibinfo{author}{Ackermann, M.} (\bibinfo{year}{2020}). \bibinfo{title}{Short-range interactions govern the dynamics and functions of microbial communities}.
\newblock \bibinfo{journal}{Nature Ecology \& Evolution} \emph{\bibinfo{volume}{4}}, \bibinfo{pages}{366--375}. \DOIprefix\doi{10.1038/s41559-019-1080-2}.
\bibitem[{Thiermann et~al.(2024)Thiermann, Sandler, Ahir, Sauls, Schroeder, Brown, Le~Treut, Si, Li, Wang and Jun}]{thiermann_tools_2024}
\bibinfo{author}{Thiermann, R.}, \bibinfo{author}{Sandler, M.}, \bibinfo{author}{Ahir, G.}, \bibinfo{author}{Sauls, J.T.}, \bibinfo{author}{Schroeder, J.}, \bibinfo{author}{Brown, S.}, \bibinfo{author}{Le~Treut, G.}, \bibinfo{author}{Si, F.}, \bibinfo{author}{Li, D.}, \bibinfo{author}{Wang, J.D.}, and \bibinfo{author}{Jun, S.} (\bibinfo{year}{2024}). \bibinfo{title}{Tools and methods for high-throughput single-cell imaging with the mother machine}.
\newblock \bibinfo{journal}{eLife} \emph{\bibinfo{volume}{12}}, \bibinfo{pages}{RP88463}. \DOIprefix\doi{10.7554/eLife.88463}.
\bibitem[{Lugagne et~al.(2024)Lugagne, Blassick and Dunlop}]{lugagne_deep_2024}
\bibinfo{author}{Lugagne, J.B.}, \bibinfo{author}{Blassick, C.M.}, and \bibinfo{author}{Dunlop, M.J.} (\bibinfo{year}{2024}). \bibinfo{title}{Deep model predictive control of gene expression in thousands of single cells}.
\newblock \bibinfo{journal}{Nature Communications} \emph{\bibinfo{volume}{15}}. \DOIprefix\doi{10.1038/s41467-024-46361-1}.
\bibitem[{Merrin(2019)}]{merrin_frontiers_2019}
\bibinfo{author}{Merrin, J.} (\bibinfo{year}{2019}). \bibinfo{title}{Frontiers in {Microfluidics}, a {Teaching} {Resource} {Review}}.
\newblock \bibinfo{journal}{Bioengineering} \emph{\bibinfo{volume}{6}}, \bibinfo{pages}{109}. \DOIprefix\doi{10.3390/bioengineering6040109}.
\bibitem[{Long et~al.(2013)Long, Nugent, Javer, Cicuta, Sclavi, Lagomarsino and Dorfman}]{long_microfluidic_2013}
\bibinfo{author}{Long, Z.}, \bibinfo{author}{Nugent, E.}, \bibinfo{author}{Javer, A.}, \bibinfo{author}{Cicuta, P.}, \bibinfo{author}{Sclavi, B.}, \bibinfo{author}{Lagomarsino, M.C.}, and \bibinfo{author}{Dorfman, K.D.} (\bibinfo{year}{2013}). \bibinfo{title}{Microfluidic chemostat for measuring single cell dynamics in bacteria}.
\newblock \bibinfo{journal}{Lab on a Chip} \emph{\bibinfo{volume}{13}}, \bibinfo{pages}{947--954}. \DOIprefix\doi{10.1039/C2LC41196B}.
\bibitem[{Shi et~al.(2024)Shi, Tabet, Milkie, Daugird, Yang, Ritter, Giovannucci and Legant}]{shi_smartllsm_2024}
\bibinfo{author}{Shi, Y.}, \bibinfo{author}{Tabet, J.S.}, \bibinfo{author}{Milkie, D.E.}, \bibinfo{author}{Daugird, T.A.}, \bibinfo{author}{Yang, C.Q.}, \bibinfo{author}{Ritter, A.T.}, \bibinfo{author}{Giovannucci, A.}, and \bibinfo{author}{Legant, W.R.} (\bibinfo{year}{2024}). \bibinfo{title}{Smart lattice light-sheet microscopy for imaging rare and complex cellular events}.
\newblock \bibinfo{journal}{Nature Methods}. \DOIprefix\doi{10.1038/s41592-023-02126-0}.
\bibitem[{Waithe et~al.(2020)Waithe, Brown, Reglinski, Diez-Sevilla, Roberts and Eggeling}]{waithe_object_2020}
\bibinfo{author}{Waithe, D.}, \bibinfo{author}{Brown, J.M.}, \bibinfo{author}{Reglinski, K.}, \bibinfo{author}{Diez-Sevilla, I.}, \bibinfo{author}{Roberts, D.}, and \bibinfo{author}{Eggeling, C.} (\bibinfo{year}{2020}). \bibinfo{title}{Object detection networks and augmented reality for cellular detection in fluorescence microscopy}.
\newblock \bibinfo{journal}{Journal of Cell Biology} \emph{\bibinfo{volume}{219}}. \DOIprefix\doi{10.1083/jcb.201903166}.
\bibitem[{Al-Hamadani et~al.(2025)Al-Hamadani, Poroszlay, Szeman-Nagy, Hajdu, Hadjidemetriou, Ferrarini and Harangi}]{alhamadani_yolo_deepsort_2025}
\bibinfo{author}{Al-Hamadani, M.N.A.}, \bibinfo{author}{Poroszlay, R.}, \bibinfo{author}{Szeman-Nagy, G.}, \bibinfo{author}{Hajdu, A.}, \bibinfo{author}{Hadjidemetriou, S.}, \bibinfo{author}{Ferrarini, L.}, and \bibinfo{author}{Harangi, B.} (\bibinfo{year}{2025}). \bibinfo{title}{Improving cell detection and tracking in microscopy images using {YOLO} and an enhanced {DeepSORT} algorithm}.
\newblock \bibinfo{journal}{Sensors} \emph{\bibinfo{volume}{25}}, \bibinfo{pages}{4361}. \DOIprefix\doi{10.3390/s25144361}.
\bibitem[{Jocher et~al.(2023)Jocher, Qiu and Chaurasia}]{jocher_ultralytics_2023}
\bibinfo{author}{Jocher, G.}, \bibinfo{author}{Qiu, J.}, and \bibinfo{author}{Chaurasia, A.} (\bibinfo{year}{2023}).
\newblock \bibinfo{title}{{Ultralytics YOLO}}. \bibinfo{publisher}{GitHub}.
\newblock \URLprefix \url{https://github.com/ultralytics/ultralytics}. \DOIprefix\doi{10.5281/zenodo.7347926}.
\bibitem[{Chiron et~al.(2022)Chiron, Le~Bec, Cordier, Pouzet, Milunov, Banderas, Di~Meglio, Sorre and Hersen}]{chiron_cybersco_2022}
\bibinfo{author}{Chiron, L.}, \bibinfo{author}{Le~Bec, M.}, \bibinfo{author}{Cordier, C.}, \bibinfo{author}{Pouzet, S.}, \bibinfo{author}{Milunov, D.}, \bibinfo{author}{Banderas, A.}, \bibinfo{author}{Di~Meglio, J.M.}, \bibinfo{author}{Sorre, B.}, and \bibinfo{author}{Hersen, P.} (\bibinfo{year}{2022}). \bibinfo{title}{{CyberSco.Py} an open-source software for event-based, conditional microscopy}.
\newblock \bibinfo{journal}{Scientific Reports} \emph{\bibinfo{volume}{12}}, \bibinfo{pages}{11579}. \DOIprefix\doi{10.1038/s41598-022-15207-5}.
\bibitem[{Passmore et~al.(2025)Passmore, Rates, Schr\"oder, van Laarhoven, Hellebrekers, van Hoef, Geurts, van Straaten, Nijenhuis, Berger, Smith, Smal and Kapitein}]{passmore_closedloop_2025}
\bibinfo{author}{Passmore, J.B.}, \bibinfo{author}{Rates, A.}, \bibinfo{author}{Schr\"oder, J.}, \bibinfo{author}{van Laarhoven, M.T.P.}, \bibinfo{author}{Hellebrekers, V.J.W.}, \bibinfo{author}{van Hoef, H.G.}, \bibinfo{author}{Geurts, A.J.M.}, \bibinfo{author}{van Straaten, W.}, \bibinfo{author}{Nijenhuis, W.}, \bibinfo{author}{Berger, F.}, \bibinfo{author}{Smith, C.S.}, \bibinfo{author}{Smal, I.}, and \bibinfo{author}{Kapitein, L.C.} (\bibinfo{year}{2025}). \bibinfo{title}{Closed-loop optogenetic control of cell biology enables outcome-driven microscopy}.
\newblock \bibinfo{journal}{Nature Communications} \emph{\bibinfo{volume}{17}}, \bibinfo{pages}{1087}. \DOIprefix\doi{10.1038/s41467-025-67848-5}.
\bibitem[{Gruenberger et~al.(2013)Gruenberger, Probst, Heyer, Wiechert, Frunzke and Kohlheyer}]{Gruenberger2013}
\bibinfo{author}{Gruenberger, A.}, \bibinfo{author}{Probst, C.}, \bibinfo{author}{Heyer, A.}, \bibinfo{author}{Wiechert, W.}, \bibinfo{author}{Frunzke, J.}, and \bibinfo{author}{Kohlheyer, D.} (\bibinfo{year}{2013}). \bibinfo{title}{{Microfluidic picoliter bioreactor for microbial single-cell analysis: Fabrication, system setup, and operation}}.
\newblock \bibinfo{journal}{Journal of Visualized Experiments}. \URLprefix \url{https://app.jove.com/t/50560}. \DOIprefix\doi{10.3791/50560}.
\bibitem[{Ravi et~al.(2024)Ravi, Gabeur, Hu, Hu, Ryali, Ma, Khedr, R\"adle, Rolland, Gustafson, Mintun, Pan, Alwala, Carion, Wu, Girshick, Doll\'ar and Feichtenhofer}]{ravi_sam2_2024}
\bibinfo{author}{Ravi, N.}, \bibinfo{author}{Gabeur, V.}, \bibinfo{author}{Hu, Y.T.}, \bibinfo{author}{Hu, R.}, \bibinfo{author}{Ryali, C.}, \bibinfo{author}{Ma, T.}, \bibinfo{author}{Khedr, H.}, \bibinfo{author}{R\"adle, R.}, \bibinfo{author}{Rolland, C.}, \bibinfo{author}{Gustafson, L.}, \bibinfo{author}{Mintun, E.}, \bibinfo{author}{Pan, J.}, \bibinfo{author}{Alwala, K.V.}, \bibinfo{author}{Carion, N.}, \bibinfo{author}{Wu, C.Y.}, \bibinfo{author}{Girshick, R.}, \bibinfo{author}{Doll\'ar, P.}, and \bibinfo{author}{Feichtenhofer, C.} (\bibinfo{year}{2024}). \bibinfo{title}{{SAM} 2: {Segment} {Anything} in {Images} and {Videos}}.
\newblock \bibinfo{journal}{arXiv}. \DOIprefix\doi{10.48550/arXiv.2408.00714}.
\bibitem[{Riba et~al.(2020)Riba, Mishkin, Ponsa, Rublee and Bradski}]{riba_kornia_2020}
\bibinfo{author}{Riba, E.}, \bibinfo{author}{Mishkin, D.}, \bibinfo{author}{Ponsa, D.}, \bibinfo{author}{Rublee, E.}, and \bibinfo{author}{Bradski, G.} (\bibinfo{year}{2020}). \bibinfo{title}{Kornia: an open source differentiable computer vision library for {PyTorch}}.
\newblock In \bibinfo{booktitle}{2020 IEEE Winter Conference on Applications of Computer Vision (WACV)}. pp. \bibinfo{pages}{3663--3672}.
\newblock \DOIprefix\doi{10.1109/WACV45572.2020.9093363}.
\bibitem[{Bradski(2000)}]{bradski_opencv_2000}
\bibinfo{author}{Bradski, G.} (\bibinfo{year}{2000}). \bibinfo{title}{The {OpenCV} library}.
\newblock \bibinfo{journal}{Dr. Dobb's Journal of Software Tools}.
\bibitem[{Gillies et~al.(2007)Gillies, van~der Wel, Van~den Bossche, Taves, Arnott, Ward et~al.}]{gillies_shapely_2007}
\bibinfo{author}{Gillies, S.}, \bibinfo{author}{van~der Wel, C.}, \bibinfo{author}{Van~den Bossche, J.}, \bibinfo{author}{Taves, M.W.}, \bibinfo{author}{Arnott, J.}, \bibinfo{author}{Ward, B.C.} et~al. (\bibinfo{year}{2007}).
\newblock \bibinfo{title}{{Shapely}}. \bibinfo{publisher}{GitHub}.
\newblock \URLprefix \url{https://github.com/shapely/shapely}. \DOIprefix\doi{10.5281/zenodo.5597138}.
\bibitem[{Gillies et~al.(2013)}]{gillies_rasterio_2013}
\bibinfo{author}{Gillies, S.} et~al. (\bibinfo{year}{2013}).
\newblock \bibinfo{title}{{Rasterio}: geospatial raster i/o for {Python} programmers}. \bibinfo{publisher}{GitHub}.
\newblock \URLprefix \url{https://github.com/rasterio/rasterio}.
\bibitem[{Zwietering et~al.(1990)Zwietering, Jongenburger, Rombouts and van~'t Riet}]{zwietering_modeling_1990}
\bibinfo{author}{Zwietering, M.H.}, \bibinfo{author}{Jongenburger, I.}, \bibinfo{author}{Rombouts, F.M.}, and \bibinfo{author}{van~'t Riet, K.} (\bibinfo{year}{1990}). \bibinfo{title}{Modeling of the bacterial growth curve}.
\newblock \bibinfo{journal}{Applied and Environmental Microbiology} \emph{\bibinfo{volume}{56}}, \bibinfo{pages}{1875--1881}. \DOIprefix\doi{10.1128/aem.56.6.1875-1881.1990}.
\bibitem[{Virtanen et~al.(2020)Virtanen, Gommers, Oliphant, Haberland, Reddy, Cournapeau, Burovski, Peterson, Weckesser, Bright, van~der Walt, Brett, Wilson, Millman, Mayorov, Nelson, Jones, Kern, Larson, Carey, Polat, Feng, Moore, VanderPlas, Laxalde, Perktold, Cimrman, Henriksen, Quintero, Harris, Archibald, Ribeiro, Pedregosa, van Mulbregt and {SciPy 1.0 Contributors}}]{virtanen_scipy_2020}
\bibinfo{author}{Virtanen, P.}, \bibinfo{author}{Gommers, R.}, \bibinfo{author}{Oliphant, T.E.}, \bibinfo{author}{Haberland, M.}, \bibinfo{author}{Reddy, T.}, \bibinfo{author}{Cournapeau, D.}, \bibinfo{author}{Burovski, E.}, \bibinfo{author}{Peterson, P.}, \bibinfo{author}{Weckesser, W.}, \bibinfo{author}{Bright, J.}, \bibinfo{author}{van~der Walt, S.J.}, \bibinfo{author}{Brett, M.}, \bibinfo{author}{Wilson, J.}, \bibinfo{author}{Millman, K.J.}, \bibinfo{author}{Mayorov, N.}, \bibinfo{author}{Nelson, A.R.J.}, \bibinfo{author}{Jones, E.}, \bibinfo{author}{Kern, R.}, \bibinfo{author}{Larson, E.}, \bibinfo{author}{Carey, C.J.}, \bibinfo{author}{Polat, {\.I}.}, \bibinfo{author}{Feng, Y.}, \bibinfo{author}{Moore, E.W.}, \bibinfo{author}{VanderPlas, J.}, \bibinfo{author}{Laxalde, D.}, \bibinfo{author}{Perktold, J.}, \bibinfo{author}{Cimrman, R.}, \bibinfo{author}{Henriksen, I.}, \bibinfo{author}{Quintero, E.A.}, \bibinfo{author}{Harris, C.R.}, \bibinfo{author}{Archibald, A.M.}, \bibinfo{author}{Ribeiro, A.H.},
  \bibinfo{author}{Pedregosa, F.}, \bibinfo{author}{van Mulbregt, P.}, and \bibinfo{author}{{SciPy 1.0 Contributors}} (\bibinfo{year}{2020}). \bibinfo{title}{{SciPy 1.0}: fundamental algorithms for scientific computing in {Python}}.
\newblock \bibinfo{journal}{Nature Methods} \emph{\bibinfo{volume}{17}}, \bibinfo{pages}{261--272}. \DOIprefix\doi{10.1038/s41592-019-0686-2}.

\end{thebibliography}


\begin{thebibliography}{4}

\providecommand{\natexlab}[1]{#1}
\providecommand{\url}[1]{\texttt{#1}}
\providecommand{\href}[2]{#2}
\providecommand{\path}[1]{#1}
\providecommand{\DOIprefix}{doi: }
\providecommand{\ArXivprefix}{arXiv: }
\providecommand{\URLprefix}{URL: }
\providecommand{\Pubmedprefix}{pmid: }
\providecommand{\doi}[1]{\href{http://dx.doi.org/#1}{\path{#1}}}
\providecommand{\Pubmed}[1]{\href{pmid:#1}{\path{#1}}}
\providecommand{\BIBand}{and}
\providecommand{\bibinfo}[2]{#2}
\ifx\xfnm\undefined \def\xfnm[#1]{\unskip,\space#1}\fi
\makeatletter\def\@biblabel#1{S#1.}\makeatother

\bibitem[{Khanam and Hussain(2024{\natexlab{a}})}]{khanam_yolov5_2024}
\bibinfo{author}{Khanam, R.}, and \bibinfo{author}{Hussain, M.} (\bibinfo{year}{2024}{\natexlab{a}}). \bibinfo{title}{What is {YOLOv5}: {A} deep look into the internal features of the popular object detector}.
\newblock \bibinfo{journal}{arXiv}. \DOIprefix\doi{10.48550/arXiv.2407.20892}.

\bibitem[{Yaseen(2024)}]{yaseen_yolov8_2024}
\bibinfo{author}{Yaseen, M.} (\bibinfo{year}{2024}). \bibinfo{title}{What is {YOLOv8}: {An} in-depth exploration of the internal features of the next-generation object detector}.
\newblock \bibinfo{journal}{arXiv}. \DOIprefix\doi{10.48550/arXiv.2408.15857}.

\bibitem[{Khanam and Hussain(2024{\natexlab{b}})}]{khanam_yolov11_2024}
\bibinfo{author}{Khanam, R.}, and \bibinfo{author}{Hussain, M.} (\bibinfo{year}{2024}{\natexlab{b}}). \bibinfo{title}{{YOLOv11}: {An} overview of the key architectural enhancements}.
\newblock \bibinfo{journal}{arXiv}. \DOIprefix\doi{10.48550/arXiv.2410.17725}.

\bibitem[{Sapkota et~al.(2025)Sapkota, Cheppally, Sharda and Karkee}]{sapkota_yolo26_2025}
\bibinfo{author}{Sapkota, R.}, \bibinfo{author}{Cheppally, R.H.}, \bibinfo{author}{Sharda, A.}, and \bibinfo{author}{Karkee, M.} (\bibinfo{year}{2025}). \bibinfo{title}{{YOLO26}: {Key} architectural enhancements and performance benchmarking for real-time object detection}.
\newblock \bibinfo{journal}{arXiv}. \DOIprefix\doi{10.48550/arXiv.2509.25164}.

\end{thebibliography}

\bigskip

\newpage

\setcounter{page}{1}
\renewcommand{\thepage}{S.\arabic{page}}

\setcounter{equation}{0}
\renewcommand{\theequation}{S.\arabic{equation}}

\section*{Supplemental information}

\bigskip

\maketitle

\renewcommand{\thesection}{S\arabic{section}}
\renewcommand{\thesubsection}{S\arabic{subsection}}
\renewcommand{\thefigure}{\textbf{S\arabic{figure}}}
\renewcommand{\thetable}{\textbf{S\arabic{table}}}
\renewcommand{\bibsection}{\section*{Supplemental references}}

\setcounter{figure}{0}
\setcounter{table}{0}

\vfill 

\footnotesize

\section*{Contents}
\noindent \textbf{Text S1.} \gls{yolo} model benchmark screening. 
\vspace{0.5em}

\noindent \textbf{Text S2.} Affine transformation for coarse alignment. 
\vspace{0.5em}

\noindent \textbf{Text S3.} Live-cell experiment analysis. 
\vspace{0.5em}

\noindent \textbf{Text S4.} Acronyms. 
\vspace{0.5em}

\noindent \textbf{Figure S1.} Exemple training batch. 
\vspace{0.5em}

\noindent \textbf{Figure S2.} Marker center detection error and speed across \gls{yolo} model configurations. 
\vspace{0.5em}

\noindent \textbf{Figure S3.} Validation of coarse alignment localization accuracy.   
\vspace{0.5em}

\noindent \textbf{Figure S4.} Automated and design-aware \gls{roi} masking for the \gls{sak} chip design. 
\vspace{0.5em}

\noindent \textbf{Table S1.} \gls{sak} \gls{roi} dimensions. 
\vspace{0.5em}

\noindent \textbf{Table S2.} Marker detection dataset statistics. 
\vspace{0.5em}

\noindent \textbf{Table S3.} Data augmentation configuration for \gls{yolo} training. 
\vspace{0.5em}

\noindent \textbf{Table S4.} \gls{yolo} model comparison on the DART test set across \gls{yolo} versions, tasks, model sizes, and input resolutions. 
\vspace{0.5em}

\noindent \textbf{Table S5.} Quantitative results of the automated \gls{dart} pipeline applied to the \gls{cglut} \gls{mlci} experiment on the \gls{sak} chip. 
\vspace{0.5em}

\noindent \textbf{Supplemental references}

\clearpage

\setcounter{subsection}{0} 

\subsection{YOLO model benchmark screening results} 
\label{sec:SI_yolo_benchmark}

\begin{figure}[H]
    \centering
    \includegraphics[width=1\linewidth]{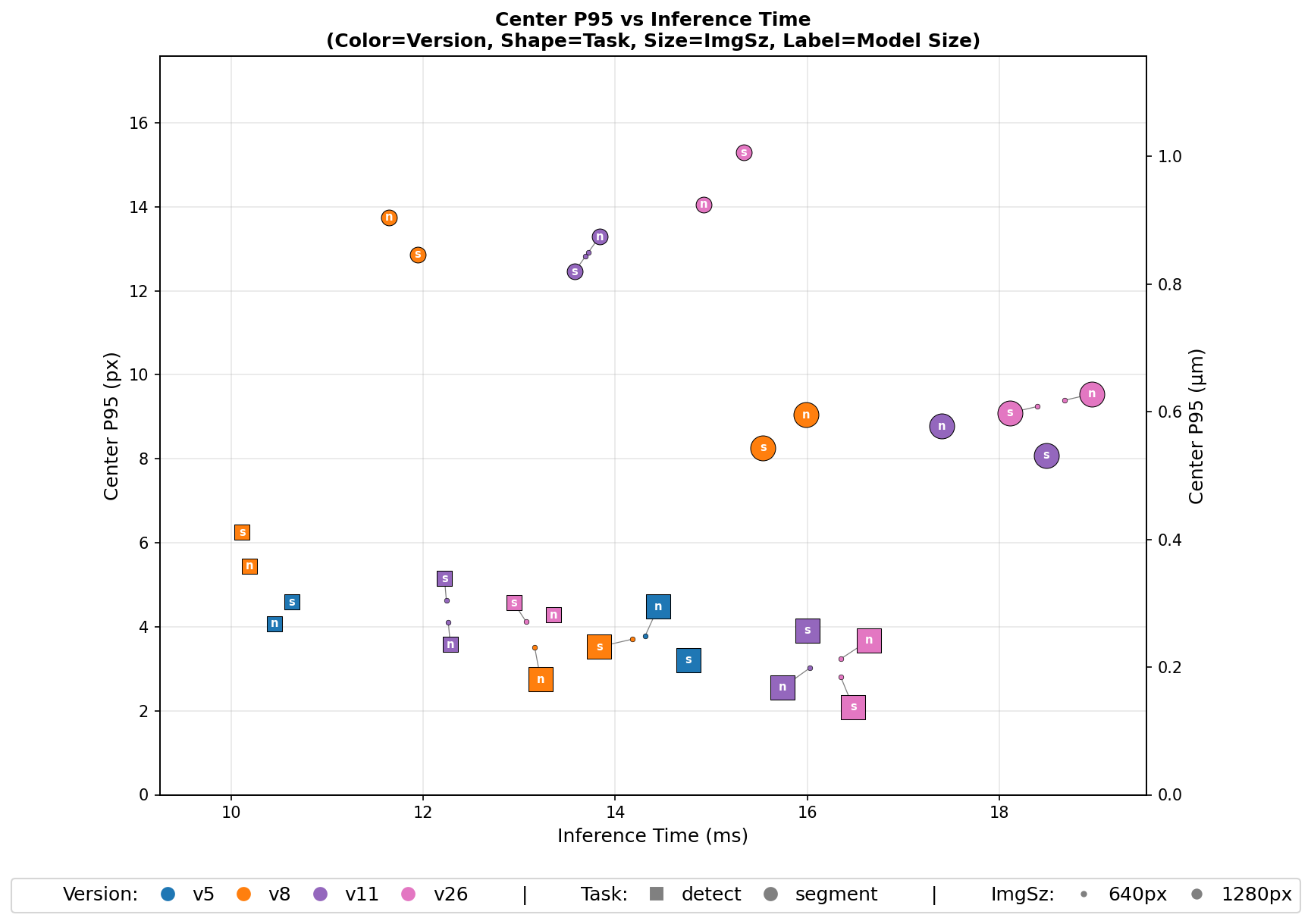}
    \caption{\textbf{Marker-center detection error and inference time across \gls{yolo} model configurations.} Marker-center detection error is reported in pixels (left axis) and micrometers (right axis), together with the inference time for a single image. Results are shown for \gls{yolo} model configurations spanning major version, training task, input resolution, and model size (n, s). Where data points lie very close together, markers are offset for readability and connected to the true measurement positions.}
    \label{SIfig:benchmark_marker_center_error_and_speed}
\end{figure}

To select a fiducial marker detector suitable for the \gls{dart} pipeline, we benchmarked \gls{yolo} architectures across model versions (v5~\cite{khanam_yolov5_2024}, v8~\cite{yaseen_yolov8_2024}, v11~\cite{khanam_yolov11_2024}, and v26~\cite{sapkota_yolo26_2025}), model sizes (n and s), input resolutions (\num{640} and \SI{1280}{\pixel}), and task types (bounding-box detection and segmentation). Marker-center precision was evaluated as the Euclidean distance between predicted and annotated marker-center positions. Model selection was based on the trade-off between marker-center accuracy and inference speed. Full per-configuration benchmark results are provided in \Cref{SItab:yolo_results} and visually summarized in \Cref{SIfig:benchmark_marker_center_error_and_speed}.

\clearpage
\subsection{Affine transformation for coarse alignment}
\label{sec:SI_map_creation}

To align the blueprint with the microscope stage coordinate system, we compute an affine transformation $T$ from three pairs of corresponding reference positions. In homogeneous coordinates, the transformation of a point $\mathbf{p}=(x_1, y_1, 1)^T$ is given by

\begin{align}
    T(\mathbf{p}) = \mathbf{A} \cdot \mathbf{p} =
    \underbrace{
    \begin{bmatrix}
        a_{11} & a_{12} & t_{1} \\
        a_{21} & a_{22} & t_{2}\\
        0 & 0 & 1
        \end{bmatrix}}_{\mathbf{A} \in \mathbb{R}^{3\times 3}}
        \cdot
        \underbrace{\begin{bmatrix}
        x_1 \\ y_1 \\ 1
    \end{bmatrix}}_{\mathbf{p} \in \mathbb{R}^3},
\end{align}
where $a_{11}, a_{12}, a_{21}, a_{22}$ encode rotation, scaling, and shear, and $t_1, t_2$ encode translation.

Given three corresponding reference positions in the blueprint coordinate system, $\mathbf{X}_{\text{blueprint}}$, and the microscope stage coordinate system, $\mathbf{X}_{\text{stage}}$
\begin{equation}
    \mathbf{X}_{\text{blueprint}} = \begin{bmatrix}
        x_1 & x_2 & x_3 \\
        y_1 & y_2 & y_3 \\
        1 & 1 & 1
    \end{bmatrix} \quad \mbox{and} \quad
    \mathbf{X}_{\text{stage}} = \begin{bmatrix}
        x_1' & x_2' & x_3'\\
        y_1' & y_2' & y_3'\\
        1 & 1 & 1
    \end{bmatrix}
\end{equation}
the transformation matrix is obtained as
\begin{equation}
    \mathbf{A} = \mathbf{X}_{\text{stage}} \cdot \mathbf{X}_{\text{blueprint}}^{-1}.
\end{equation}
The inverse exists when the three reference positions are non-collinear, that is, when $\mathbf{X}_{\text{blueprint}}$ has full rank. In practice, reference positions were selected near the edges or corners of the chip to maximize spatial coverage of the resulting transformation.

\clearpage
\subsection{Live-cell experiment analysis}
\label{sec:SI_live_cell_experiment}

Across the full live-cell validation dataset, the fully automated \gls{dart} image-analysis pipeline processed \num{1739} images in \SI{31}{\minute} and identified more than \num{500000} individual cells while removing interior microfluidic structures. This corresponds to a throughput of approximately \num{3300} images per hour and \num{960000} segmented cells per hour on the benchmarking hardware described in the Methods, without application-specific optimization of the segmentation model. Relative to the experimental imaging time of \SI{540}{\minute}, this yields a real-time factor of \num{17.4}.

\clearpage
\subsection{Acronyms}
\printglossary[type=\acronymtype,nonumberlist,title={}]

\clearpage
\begin{figure}[H]
    \centering
    \includegraphics[width=0.75\linewidth]{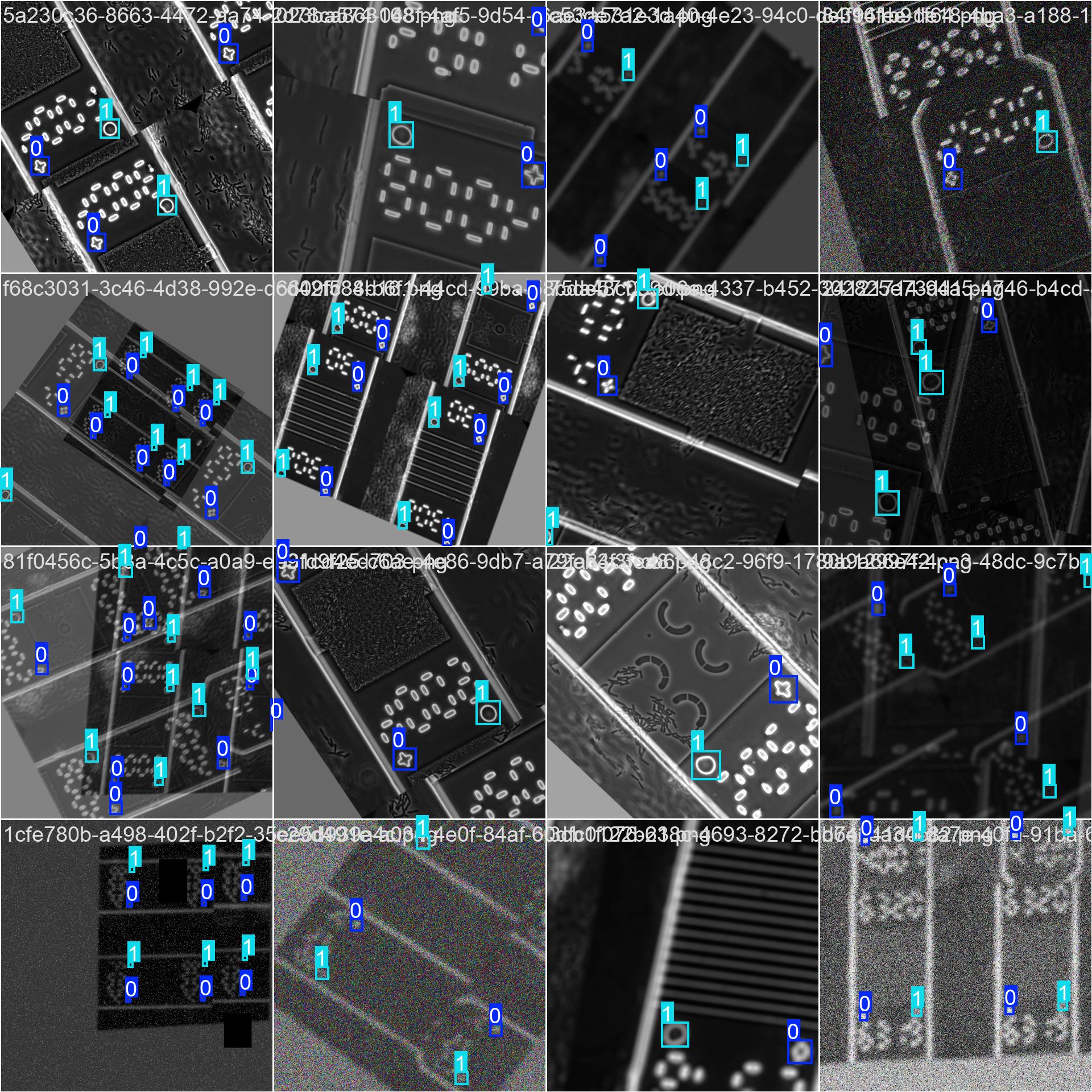}
    \caption{\textbf{Example training batch.} Batch of \num{16} augmented images illustrating simulated suboptimal microscopy conditions. Annotated bounding boxes indicate cross (\bluerect) and circle (\tealrect) fiducial markers.}
    \label{SIfig:train-batch}
\end{figure}

\clearpage

\clearpage
\begin{figure}[H]
    \centering
    \includegraphics[width=.8\linewidth]{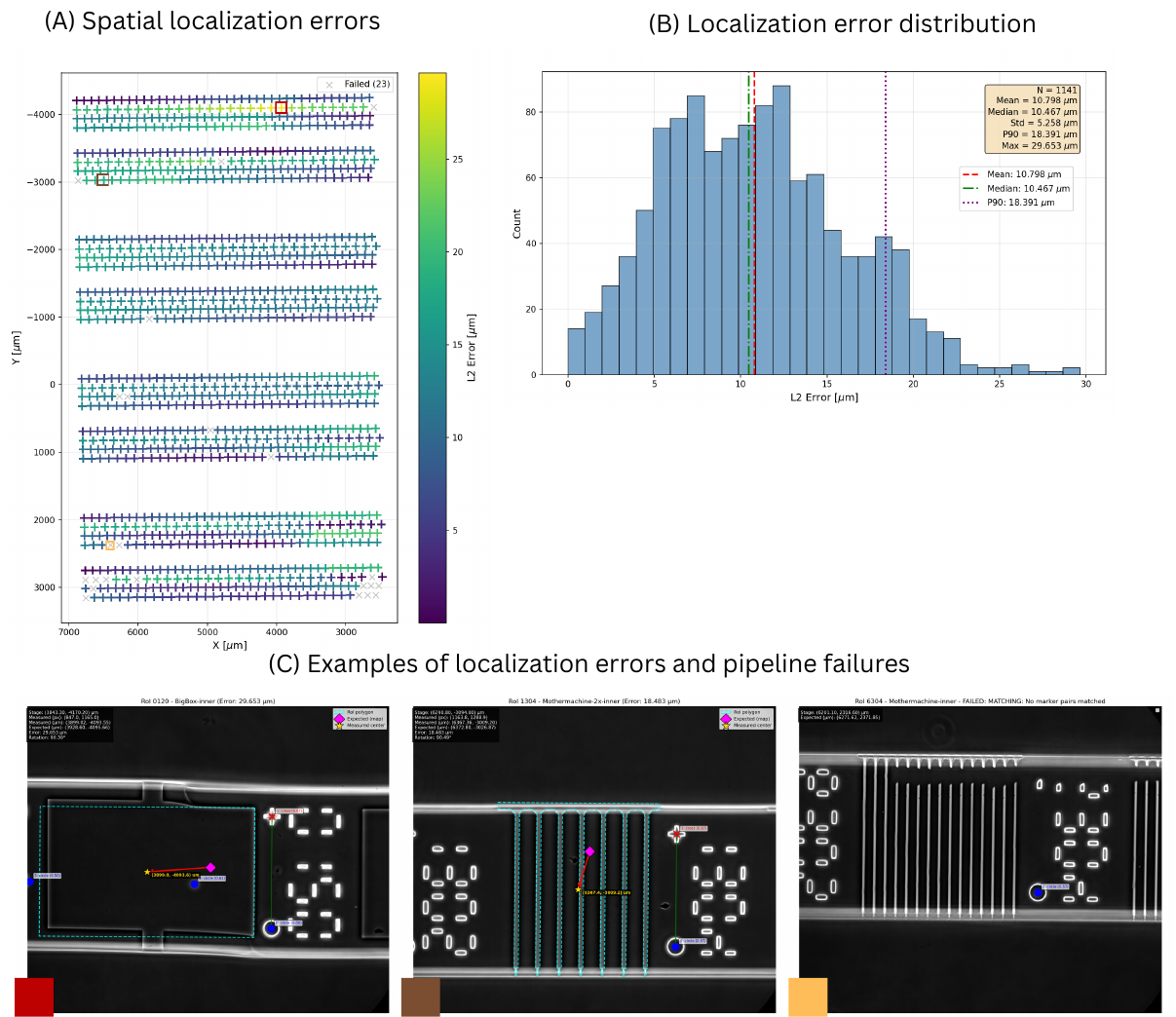}
    \caption{\textbf{Validation of coarse alignment localization accuracy.} (A)~Spatial map of localization errors across the microfluidic chip. Each cross denotes an \gls{roi} position, and the color indicates the absolute position error between the map-predicted and image-derived \gls{roi} center positions. Grey ``x'' symbols denote preprocessing failures. (B)~Distribution of the same absolute position errors. The median (\SI{10.46}{\micro\metre}) and \num{90}th percentile (\SI{18.39}{\micro\metre}) are indicated by red and purple dashed lines, respectively. (C)~Representative examples of localization errors and pipeline failures for three \glspl{roi}. The colored box indicates the corresponding physical position, which is also highlighted in (A). The purple diamond marks the \gls{roi} center predicted by coarse alignment, and the yellow star marks the image-derived \gls{roi} center based on detected fiducial markers. From left to right, the examples illustrate deformed microfluidic structures, fine mother-machine structures, and fabrication defects.}
    \label{SIfig:dmc_map_validation}
\end{figure}

\clearpage
\begin{figure}[H]
    \centering
    \includegraphics[width=.9\linewidth]{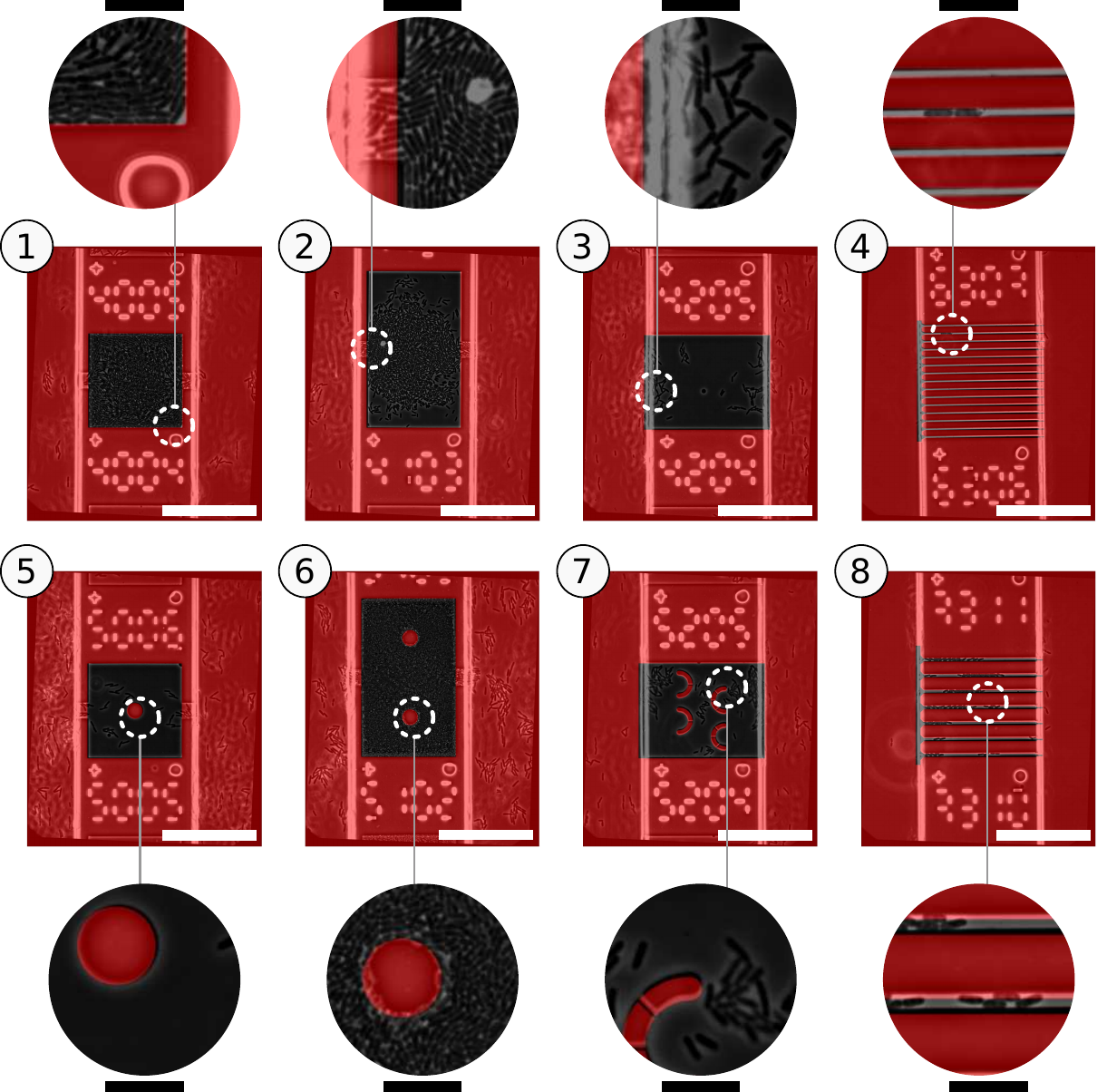}
    \caption{\textbf{Automated and design-aware \gls{roi} masking for the \gls{sak} chip design.} Panels 1--8 show representative phase-contrast images of the eight different \gls{roi} designs. The blueprint-derived mask is shown in red and aligned to each image using the \gls{dart} pipeline. For each image, the white circle indicates the location of a \num{5}-fold magnified inset highlighting the local masking accuracy. Black bars: \SI{10}{\micro\metre} (insets); white bars: \SI{60}{\micro\metre} (full images).}   
    \label{SIfig:dmc_rois_masked}
\end{figure}

\clearpage
\begin{table}[H]
    \centering
    \caption{\textbf{\gls{sak} \gls{roi} dimensions.} Physical dimensions of the eight different chamber structure designs.\\}
    \label{SItab:roi_dimensions}
    \begin{tabular}{l r}
        \toprule
        Chamber Type & Dimensions [\SI{}{\micro\metre\squared}]\\
        \midrule
        NormaleBox-inner (1) & \num{60}~$\times$~\num{60}  \\
        BigBox-inner (2) & \num{60}~$\times$~\num{100} \\
        OpenBox-inner (3) & \num{60}~$\times$~\num{80} \\
        Mothermachine-1x-inner (4) & \num{15}~$\times$~\num{1}~$\times$~\num{80} \\
        NormaleBox-pillar-inner (5) & \num{60}~$\times$~\num{60} \\
        BigBox-pillar-inner (6) & \num{60}~$\times$~\num{100} \\
        OpenBox-collector-inner (7) & \num{60}~$\times$~\num{80} \\
        Mothermachine-2x-inner (8) & \num{7}~$\times$~\num{2}~$\times$~\num{80} \\
        \bottomrule
    \end{tabular}
\end{table}

\clearpage
\begin{table}[H]
    \centering
    \caption{\textbf{Marker detection dataset statistics.} 
    Summary of image, cell, and marker counts for the training, validation, and test splits of the annotated fiducial-marker dataset. Training and validation data were obtained by an \num{80}/\num{20} split of the annotated development dataset, while the test set was collected independently and used for model selection.\\}
    \label{SItab:dataset-statistics}
    \begin{tabular}{lrrrr}
        \toprule
        Split & Images & Total Objects & Cross & Circle \\
        \midrule
        Training & \num{192} & \num{584} & \num{292} & \num{292} \\
        Validation   &  \num{47} & \num{130} &  \num{66} &  \num{64} \\
        Test  &  \num{80} & \num{228} & \num{114} & \num{114} \\
        \midrule
        \textbf{Total} & \textbf{\num{319}} & \textbf{\num{942}} & \textbf{\num{472}} & \textbf{\num{470}} \\
        \bottomrule
    \end{tabular}
\end{table}

\clearpage
\begin{table}[H]
    \centering
    \caption{\textbf{Data augmentation configuration for \gls{yolo} training.} The table lists the augmentation operations used during training. ``OneOf'' denotes a random selection of one transform from the listed options.\\}
    \label{SItab:augmentations}
    \begin{tabular}{llll}
        \toprule
        \textbf{Category} & \textbf{Augmentation} & \textbf{Parameters} & \textbf{Probability} \\
        \midrule
        \multicolumn{4}{l}{\textit{Built-in \gls{yolo} augmentations}} \\
        \midrule
        \multirow{5}{*}{Geometric}
            & Rotation & $\pm$\SI{90}{\degree} & \num{1.0} \\
            & Vertical Flip & -- & \num{0.5} \\
            & Scale & $\pm$\SI{70}{\degree} & \num{1.0} \\
            & Shear & $\pm$\SI{5}{\degree} & \num{1.0} \\
            & Perspective & \num{0.0} & -- \\
        \midrule
        \multirow{3}{*}{Color (HSV)}
            & Hue shift & $\pm$ \num{0.02} & \num{1.0} \\
            & Saturation shift & $\pm$ \num{0.7} & \num{1.0} \\
            & Value shift & $\pm$ \num{0.5} & \num{1.0} \\
        \midrule
        \multirow{2}{*}{Composition}
            & MixUp & $\alpha$~=~\num{0.15} & \num{0.15} \\
            & Copy-Paste & -- &\num{ 0.1} \\
        \midrule
        \multicolumn{4}{l}{\textit{Custom albumentations augmentations (microscopy-specific)}} \\
        \midrule
        \multirow{3}{*}{Blur (OneOf)}
            & Gaussian Blur & kernel: \num{3}--\num{15} & \multirow{3}{*}{\num{0.75}} \\
            & Motion Blur & kernel: \num{3}--\num{15} & \\
            & Defocus & radius: \num{3}--\num{10} & \\
        \midrule
        \multirow{3}{*}{Noise (OneOf)}
            & Gaussian Noise & $\sigma$: \num{0.02}--\num{0.15} & \multirow{3}{*}{\num{0.4}} \\
            & ISO Noise & intensity: \num{0.1}--\num{0.8} & \\
            & Multiplicative Noise & multiplier: \num{0.8}--\num{1.2} & \\
        \midrule
        \multirow{4}{*}{Illumination}
            & Brightness/Contrast & $\pm$ \num{0.2} & \num{0.5} \\
            & Gamma Correction & $\gamma$: \num{70}--\num{130} & \num{0.3} \\
            & CLAHE & clip: \num{4.0} & \num{0.1} \\
            & JPEG Compression & quality: \num{50}--\num{95} & \num{0.1} \\
        \midrule
        \multirow{2}{*}{Dropout (OneOf)}
            & Pixel Dropout & probability: \num{0.03} & \multirow{2}{*}{\num{0.3}} \\
            & Coarse Dropout & holes: \num{1}--\num{8}, size: \num{8}--\SI{32}{\pixel} & \\
        \bottomrule
    \end{tabular}
\end{table}

\clearpage
\begin{table}[H]
    \centering
    \scriptsize
    \caption{\textbf{YOLO model comparison on the DART test set across \gls{yolo} versions, tasks, model sizes, and input resolutions.} Reported metrics include mAP@0.5 and mAP@0.5:0.95 as computed by the \gls{yolo} implementation, the median marker-center detection error (Ctr Med), its \num{95}th percentile (Ctr P95), and the inference time measured at batch size \num{1}. The model selected for fiducial marker detection is highlighted in cyan.\\}
    \label{SItab:yolo_results}
    \begin{tabular}{l l l r || rrrrr}
        \toprule
        Version & Task & Size & ImgSz & mAP@0.5 $\uparrow$ & mAP@0.5:0.95 $\uparrow$ & Ctr Med (px) $\downarrow$ & Ctr P95 (px) $\downarrow$ & Infer (ms) $\downarrow$ \\
        \midrule
        v5 & detect & n & 640 & 0.982 & 0.907 & 1.86 & 4.08 & \SI[separate-uncertainty=true]{10.4(0.6)}{} \\
        v5 & detect & n & 1280 & 0.983 & 0.889 & 2.03 & 3.77 & \SI[separate-uncertainty=true]{14.3(0.4)}{} \\
        v5 & detect & s & 640 & 0.983 & 0.886 & 2.01 & 4.59 & \SI[separate-uncertainty=true]{10.6(0.8)}{} \\
        v5 & detect & s & 1280 & 0.979 & 0.889 & 1.68 & 3.21 & \SI[separate-uncertainty=true]{14.8(1.6)}{} \\
        v8 & detect & n & 640 & 0.979 & 0.866 & 2.63 & 5.44 & \SI[separate-uncertainty=true]{10.2(0.4)}{} \\
        v8 & segment & n & 640 & 0.987 & 0.907 & 7.04 & 13.75 & \SI[separate-uncertainty=true]{11.6(0.5)}{} \\
        v8 & detect & n & 1280 & 0.984 & 0.906 & 1.88 & 3.51 & \SI[separate-uncertainty=true]{13.2(1.5)}{} \\
        v8 & segment & n & 1280 & 0.985 & 0.959 & 4.50 & 9.05 & \SI[separate-uncertainty=true]{16.0(0.9)}{} \\
        v8 & detect & s & 640 & 0.986 & 0.848 & 2.99 & 6.25 & \textbf{\SI[separate-uncertainty=true]{10.1(1.1)}{}} \\
        v8 & segment & s & 640 & \textbf{0.989} & 0.922 & 7.13 & 12.86 & \SI[separate-uncertainty=true]{11.9(0.8)}{} \\
        v8 & detect & s & 1280 & 0.980 & 0.859 & 1.81 & 3.71 & \SI[separate-uncertainty=true]{14.2(0.4)}{} \\
        v8 & segment & s & 1280 & 0.988 & \textbf{0.962} & 4.86 & 8.26 & \SI[separate-uncertainty=true]{15.5(1.1)}{} \\
        v11 & detect & n & 640 & 0.976 & 0.891 & 2.00 & 4.10 & \SI[separate-uncertainty=true]{12.3(0.4)}{} \\
        v11 & segment & n & 640 & 0.972 & 0.904 & 7.41 & 12.92 & \SI[separate-uncertainty=true]{13.7(0.9)}{} \\
        v11 & detect & n & 1280 & 0.977 & 0.910 & \textbf{1.35} & 3.02 & \SI[separate-uncertainty=true]{16.0(0.4)}{} \\
        v11 & segment & n & 1280 & 0.984 & 0.955 & 4.33 & 8.77 & \SI[separate-uncertainty=true]{17.4(1.0)}{} \\
        v11 & detect & s & 640 & 0.989 & 0.879 & 2.57 & 4.62 & \SI[separate-uncertainty=true]{12.2(0.8)}{} \\
        v11 & segment & s & 640 & 0.980 & 0.904 & 7.04 & 12.82 & \SI[separate-uncertainty=true]{13.7(0.4)}{} \\
        v11 & detect & s & 1280 & 0.985 & 0.899 & 1.82 & 3.91 & \SI[separate-uncertainty=true]{16.0(0.5)}{} \\
        v11 & segment & s & 1280 & 0.986 & 0.939 & 4.80 & 8.08 & \SI[separate-uncertainty=true]{18.5(0.8)}{} \\
        v26 & detect & n & 640 & 0.957 & 0.865 & 2.07 & 4.28 & \SI[separate-uncertainty=true]{13.4(0.8)}{} \\
        v26 & segment & n & 640 & 0.910 & 0.793 & 7.48 & 14.06 & \SI[separate-uncertainty=true]{14.9(0.6)}{} \\
        v26 & detect & n & 1280 & 0.972 & 0.913 & 1.80 & 3.24 & \SI[separate-uncertainty=true]{16.4(0.3)}{} \\
        v26 & segment & n & 1280 & 0.955 & 0.929 & 4.92 & 9.39 & \SI[separate-uncertainty=true]{18.7(0.9)}{} \\
        v26 & detect & s & 640 & 0.973 & 0.888 & 2.13 & 4.12 & \SI[separate-uncertainty=true]{13.1(1.1)}{} \\
        v26 & segment & s & 640 & 0.981 & 0.921 & 9.03 & 15.29 & \SI[separate-uncertainty=true]{15.3(0.5)}{} \\
        \rowcolor{cyan!10} \textbf{v26} & \textbf{detect} & \textbf{s} & \textbf{1280} & 0.979 & 0.895 & 1.50 & \textbf{2.81} & \SI[separate-uncertainty=true]{16.4(1.0)}{} \\
        v26 & segment & s & 1280 & 0.985 & 0.960 & 5.24 & 9.24 & \SI[separate-uncertainty=true]{18.4(1.0)}{} \\
        \bottomrule
    \end{tabular}
\end{table}

\clearpage
\begin{landscape}
\begin{table}[H]
    \centering
    \scriptsize
    \caption{\textbf{Quantitative results of the automated \gls{dart} pipeline applied to the \gls{cglut} \gls{mlci} experiment on the \gls{sak} chip.} Chamber types are abbreviated as follows: (1) NormaleBox-inner, (2) BigBox-inner, (3) OpenBox-inner, (5) NormaleBox-pillar-inner, (6) BigBox-pillar-inner, (7) OpenBox-collector-inner, and (8) Mothermachine-2x-inner. Pipeline-step abbreviations are: Det.\ = marker detection, Match = marker-pair matching, Rot.\ = image rotation, Mask = masking, and Seg.\ = cell segmentation using Cellpose-SAM.\\}
    \setlength{\tabcolsep}{4pt}
    \begin{tabular}{l r r r r r r r | r}
        \toprule
        Metric & (1) & (2) & (3) & (5) & (6) & (7) & (8) & \textbf{All} \\
        \midrule
        N images                & 370             & 185               & 259             & 333             & 222               & 222             & 148            & \textbf{1739} \\
        \midrule
        Det. (ms)        & \SI[separate-uncertainty=true]{19.7(41.2)}{} & \SI[separate-uncertainty=true]{17.0(2.8)}{}    & \SI[separate-uncertainty=true]{18.3(2.5)}{}  & \SI[separate-uncertainty=true]{17.8(1.6)}{}  & \SI[separate-uncertainty=true]{18.3(6.7)}{}    & \SI[separate-uncertainty=true]{17.8(3.2)}{}  & \SI[separate-uncertainty=true]{18.0(1.7)}{} & \SI[separate-uncertainty=true]{18.3(19.3)}{} \\
    Match (ms)       & \SI[separate-uncertainty=true]{0.2(0.0)}{}   & \SI[separate-uncertainty=true]{0.2(0.0)}{}     & \SI[separate-uncertainty=true]{0.2(0.0)}{}   & \SI[separate-uncertainty=true]{0.2(0.0)}{}   & \SI[separate-uncertainty=true]{0.2(0.0)}{}     & \SI[separate-uncertainty=true]{0.2(0.0)}{}   & \SI[separate-uncertainty=true]{0.2(0.0)}{}  & \SI[separate-uncertainty=true]{0.2(0.0)}{} \\
    Rot. (ms)        & \SI[separate-uncertainty=true]{13.1(5.7)}{}  & \SI[separate-uncertainty=true]{10.7(4.3)}{}    & \SI[separate-uncertainty=true]{13.7(5.5)}{}  & \SI[separate-uncertainty=true]{12.7(5.0)}{}  & \SI[separate-uncertainty=true]{13.1(8.2)}{}    & \SI[separate-uncertainty=true]{11.7(3.0)}{}  & \SI[separate-uncertainty=true]{10.7(2.8)}{}  & \SI[separate-uncertainty=true]{12.5(5.4)}{} \\
    Mask (ms)        & \SI[separate-uncertainty=true]{8.4(3.2)}{}   & \SI[separate-uncertainty=true]{9.2(3.9)}{}     & \SI[separate-uncertainty=true]{7.6(2.3)}{}   & \SI[separate-uncertainty=true]{7.6(2.0)}{}   & \SI[separate-uncertainty=true]{10.0(3.8)}{}     & \SI[separate-uncertainty=true]{8.1(1.4)}{}   & \SI[separate-uncertainty=true]{7.1(1.3)}{}  & \SI[separate-uncertainty=true]{8.3(2.9)}{} \\
    Seg. (ms)        & \SI[separate-uncertainty=true]{889.7(489.9)}{} & \SI[separate-uncertainty=true]{1580.3(1212.5)}{} & \SI[separate-uncertainty=true]{786.1(280.4)}{} & \SI[separate-uncertainty=true]{939.9(513.5)}{} & \SI[separate-uncertainty=true]{1418.9(1012.3)}{} & \SI[separate-uncertainty=true]{909.0(316.2)}{} & \SI[separate-uncertainty=true]{894.8(57.6)}{} & \SI[separate-uncertainty=true]{1027.8(695.0)}{} \\
    \midrule
    Total (ms)       & \SI[separate-uncertainty=true]{931.1(497.6)}{} & \SI[separate-uncertainty=true]{1617.4(1214.8)}{} & \SI[separate-uncertainty=true]{825.8(281.7)}{} & \SI[separate-uncertainty=true]{978.2(514.7)}{} & \SI[separate-uncertainty=true]{1460.5(1013.3)}{} & \SI[separate-uncertainty=true]{946.7(316.5)}{} & \SI[separate-uncertainty=true]{930.8(57.9)}{} & \SI[separate-uncertainty=true]{1067.0(697.1)}{} \\
    w/o Seg (ms)     & \SI[separate-uncertainty=true]{41.4(44.7)}{} & \SI[separate-uncertainty=true]{37.2(6.7)}{}    & \SI[separate-uncertainty=true]{39.7(6.4)}{}  & \SI[separate-uncertainty=true]{38.3(6.2)}{}  & \SI[separate-uncertainty=true]{41.6(12.2)}{}    & \SI[separate-uncertainty=true]{37.7(4.9)}{}  & \SI[separate-uncertainty=true]{36.0(3.5)}{} & \SI[separate-uncertainty=true]{39.2(21.7)}{} \\
    FPS w/ Seg       & \SI[separate-uncertainty=true]{1.07(0.03)}{} & \SI[separate-uncertainty=true]{0.62(0.03)}{}   & \SI[separate-uncertainty=true]{1.21(0.03)}{} & \SI[separate-uncertainty=true]{1.02(0.03)}{} & \SI[separate-uncertainty=true]{0.68(0.03)}{}   & \SI[separate-uncertainty=true]{1.06(0.02)}{} & \SI[separate-uncertainty=true]{1.07(0.01)}{} & \SI[separate-uncertainty=true]{0.94(0.01)}{} \\
    FPS w/o Seg      & \SI[separate-uncertainty=true]{24.15(1.36)}{} & \SI[separate-uncertainty=true]{26.88(0.36)}{}  & \SI[separate-uncertainty=true]{25.19(0.25)}{} & \SI[separate-uncertainty=true]{26.11(0.23)}{} & \SI[separate-uncertainty=true]{24.04(0.47)}{}  & \SI[separate-uncertainty=true]{26.53(0.23)}{} & \SI[separate-uncertainty=true]{27.78(0.22)}{} & \SI[separate-uncertainty=true]{25.51(0.34)}{} \\
        \bottomrule
    \end{tabular}
    \label{SItab:dart_masking_pipeline_details}
\end{table}

\end{landscape}

\clearpage
\begingroup{}

\endgroup{}

\end{document}